\documentclass{aa}  

\usepackage{graphicx}
\usepackage{txfonts}
\usepackage{float}
\usepackage{placeins}
\title{A lack of LAEs within 5Mpc of a luminous quasar in an overdensity at z=6.9: potential evidence of quasar negative feedback at protocluster scales}
\titlerunning{A lack of LAEs within 5Mpc of a luminous quasar in an overdensity at z=6.9}
\begin{document}

   \author{Trystan S. Lambert
          \inst{1, 2}\fnmsep\thanks{trystanscottlambert@gmail.com}
          \and
          R. J. Assef\inst{1}
          \and
          C. Mazzucchelli\inst{1}
          \and
          E. Ba\~{n}ados \inst{3}
          \and
          M. Aravena\inst{1}
          \and
          F. Barrientos\inst{4}
          \and 
          J. Gonz\'alez-L\'opez\inst{5, 1}
          \and
          W. Hu \inst{6, 7}
          \and
          L. Infante \inst{4}
          \and
          S. Malhotra \inst{8}
          \and
          C. Moya-Sierralta \inst{4}
          \and
          J. Rhoads \inst{8}
          \and 
          F. Valdes \inst{9}
          \and
          J. Wang \inst{10}
          \and
          I. G. B. Wold \inst{8, 11, 12}
          \and
          Z. Zheng \inst{13}
          }

   \institute{Instituto de Estudios Astrof\'isicos, Facultad de Ingenier\'ia y Ciencias, Universidad Diego Portales, Av. Ej\'ercito Libertador 441, Santiago, Chile 
        \and
    ICRAR, The University of Western Australia, 35 Stirling Highway, Crawley, WA 6009, Australia
         \and
    Max Planck Institut f\"ur Astronomie, K\"onigstuhl 17, D-69117, Heidelberg, Germany
    \and
    Instituto de Astrof\'isica and Centro de Astroingenier\'ia, Facultad de F\'isica, Pontificia Universidad Cat\'olica de Chile, Casilla 306, Santiago 22, Chile
    \and 
    Las Campanas Observatory, Carnegie Institution of Washington, Casilla 601, La Serena, Chile
    \and
    Department of Physics and Astronomy, Texas A\&M University, College Station, TX 77843-4242, USA
    \and 
    George P. and Cynthia Woods Mitchell Institute for Fundamental Physics and Astronomy, Texas A\&M University, College Station, TX 77843-4242, USA
    \and
    Astrophysics Science Division, Code 665, NASA Goddard Space Flight Center, 8800 Greenbelt Rd., Greenbelt, MD 20771, USA
    \and
    NSF’s National Optical/Infrared Research Laboratory, 950 N. Cherry Ave.Tucson, AZ 85719, USA
    \and
    CAS Key Laboratory for Research in Galaxies and Cosmology, Department of Astronomy, University of Science and Technology of China, Hefei, Anhui 230026, People’s Republic of China
    \and
    Department of Physics, The Catholic University of America, Washington, DC 20064, USA
    \and
    Center for Research and Exploration in Space Science and Technology, NASA/GSFC, Greenbelt, MD 20771
    \and
    Key Laboratory for Research in Galaxies and Cosmology, Shanghai Astronomical Observatory, Chinese Academy of Sciences, 80 Nandan Road, Shanghai 200030, People’s Republic of China
             }

   \date{Received 12/02/2024; accepted 10/07/2024}

% \abstract{}{}{}{}{} 
% 5 {} token are mandatory
 
  \abstract
   {High-redshift quasars are thought to live in the densest regions of space which should be made evident by an overdensity of galaxies around them. However, campaigns to identify these overdensities through the search of Lyman Break Galaxies (LBGs) and Lyman $\alpha$ emitters (LAEs) have had mixed results. These may be explained by either the small field of view of some of the experiments, the broad redshift ranges targeted by LBG searches, and by the inherent large uncertainty of quasar redshifts estimated from UV emission lines, which makes it difficult to place the Ly-$\alpha$ emission line within a narrowband filter. Here we present a three square degree search ($\sim 1000$ pMpc$^{2}$) for LAEs around the $z=6.9$ quasar VIK~J2348--3054 using the Dark Energy CAMera (DECam), housed on the 4m Blanco telescope, finding 38 LAEs. The systemic redshift of VIK J2348--3054 is known from ALMA [CII] observations and place the Ly-$\alpha$ emission line of companions within the NB964 narrowband of DECam. This is the largest field of view LAE search around a $z>6$ quasar conducted to date. We find that this field is $\sim$ 10 times more overdense when compared to the Chandra Deep-Field South, observed previously with the same instrumental setup as well as several combined blank fields. This is strong evidence that VIK~J2348--3054 resides in an overdensity of LAEs over several Mpc. Surprisingly, we find a lack of LAEs within 5~physical~Mpc of the quasar and take this to most likely be evidence of the quasar suppressing star formation in its immediate vicinity. This result highlights the importance of performing overdensity searches over large areas to properly assess the density of those regions of the Universe.}
   
   \keywords{Cosmology: reionization --
                Galaxies: high-redshift --
                Galaxies: quasars: individual:VIK J2348--3054
               }

   \maketitle
%
%-------------------------------------------------------------------

\section{Introduction}
Several high-redshift quasars have been identified within the Epoch of Reionization \citep{Fan2006, Venemans2007, Willott2010, Venemans2013, Wang2021, EduardoReviewQSO} which is believed to have ended at $z=5.3$ \citep{Bosman2022}. In many cases, the super massive black holes (SMBHs) at the core of these quasars have masses in excess of $10^9~M_{\odot}$ \citep[e.g.,][]{Farina2022, Eilers2023, Chiara2023, Yang2023} but have had a relatively short amount of time (less than 1 Gyr) for accreting the large amounts of material needed to build up those masses \citep{Volontari2012, Inayoshi2020}. This can only be feasible if these high-redshift quasars are located within very dense regions of the early Universe \citep{Decarli2019}. Simulations have indeed shown that we should expect to find these very massive SMBHs---and by analogy, quasars---in the most massive dark matter haloes at $z>6$ and these should be traced by highly clustered galaxies \citep{Angulo2012, Costa2014}.

Observationally confirming the predictions made by simulations requires searching for galaxies around high-redshift quasars and comparing the densities found to blank fields (i.e., regions of sky without any quasars at similar redshifts that provide an estimate of the average density of the Universe). Yet this is difficult to do:  whilst we can observe the quasar at high-redshifts with relative ease, identifying the companion galaxies tracing the overdensity is a lot more difficult because of how much fainter they are.

 The main observational strategy used to search for galaxies around quasars of a known redshift is photometrically identifying either Lyman-break galaxies \citep[LBGs, e.g.,][]{Kim2009, Utsumi2010, Husband2013, Morselli2014, Dejene2023} or Lyman-$\alpha$ emitters \citep[LAEs, e.g.,][]{Eduardo2013, goto2017, Chiara2017, Ota2018} by using broadband (FWHM $\sim 1000$ \AA) photometry to identify the spectral feature created by the Lyman break and IGM absorption for the former and a combination of broad- and narrowband (FWHM $\sim 100$ \AA) photometry to identify the Lyman alpha emission line for the latter. However, results across multiple studies using these techniques to estimate the environmental densities around these high-redshift quasars have resulted in varying, and often conflicting, results. Some studies have detected overdense regions around quasars \citep{Zheng2006, Utsumi2010, Capak2011, Husband2013, Morselli2014, Balmaverde2017, Decarli2019, Mignoli2020} while others have not \citep{Stiavelli2005, Willott2005, Eduardo2013, Simpson2014, goto2017, Chiara2017}. Sometimes finding conflicting results for the same targets: e.g., SDSS J1030+0524 where \cite{Willott2005} did not find an overdensity but \cite{Stiavelli2005,Kim2009, Morselli2014, Balmaverde2017} did; likewise for SDSS J1048+4637 and SDSS J1148+5251 where \cite{Willott2005} and \cite{Kim2009} did not find overdensities but later \cite{Morselli2014} did. In order to explain these contrasting results, observational biases need to be considered. 

 For example, consider the difference in redshift resolution ($\Delta z$) between LBGs and LAEs studies. LBG searches generally have $\Delta z \sim 1$ \citep{Chiara2017} when using three broadband filters, and even though this resolution can be improved with more filters and unique set-ups \citep[e.g., see][who obtain a resolution of $\Delta z \sim 0.3$]{GarciaVergara2017}, LAE searches, because of the small wavelength range of the narrowband, have resolutions of $\Delta z \sim$ 0.1 \citep{Hu2019}, making LAEs a more precise method. This implies that galaxies that are found using the LBG technique have a higher probability of not being associated to the central source \citep{Eduardo2013}.

 Another observational effect, which is often not considered when searching for overdensities around quasars, is the method used to determine the quasar redshift. Most studies adopted quasar redshifts that have been measured using rest-frame UV lines. However, with the advent of ALMA, it is now clear that the redshifts from UV can be offset by thousands of km~s$^{-1}$ from the host galaxy's systemic redshifts \citep[e.g.,][]{Decarli2018, Schindler2020, Tanio2021}. This offset can often be large enough to shift the Ly-$\alpha$ emission outside of a narrowband filter in LAE searches, potentially resulting in false non-detections of overdensities. To date, only \cite{Ota2018}, who searched over a 0.2 square degree area, has performed a LAE search using a [CII] redshift instead of rest-frame UV. While they did not find an overdensity, it is worth noting that the Ly-$\alpha$ line in their case, was near the edge of their narrowband filter.  
 
The size of the search area may also affect the identification of overdensities. Protocluster areas are expected to be on scales of $\sim 1^{\circ}$ \citep{Overzier2009, Balmaverde2017}. \cite{Chiang2017} used a semi-analytical model anchored on the Millennium simulation and found that, on average, haloes with radii $\sim 1$ cMpc at z=0 are extended over $\sim$ 10 cMpc ( $\sim 1.25$ pMpc) at z=7; \cite{Overzier2009} suggests that protoclusters typically extend over $\sim$ 25 cMpc but can reach up to 75 cMpc. However, most studies probe areas significantly smaller than this \citep[e.g.,][]{Eduardo2013,Chiara2017}, although there has been an effort made in the submm regime, specifically by \cite{Li2023} to search for submm galaxies around these quasars over larger areas.

Accounting for these observational biases would require 1) a quasar with a known, systemic redshift, 2) a telescope and instrument with a large FoV, and 3) a narrowband filter at the correct wavelength range to capture Lyman alpha at the systemic velocity. By coincidence, this is exactly the case for the quasar VIK~2348--3054 \citep[RA = 23:48:33.34, Dec = $-$30:54:10.0;][]{Venemans2013}, which has a confirmed [CII] redshift of $z=  6.9018 \pm 0.0007$ \citep{Venemans2016}, and whose Ly-$\alpha$ emission falls within the response of the narrowband filter NB964 \citep{Zheng2019} available for the Dark Energy Camera \citep[DECam,][]{Flaugher2015} on the 4m Blanco telescope at Cerro Tololo Inter-American Observatory (CTIO), which has a 3 square degree FoV. The narrowband filter NB964 was designed for the Lyman Alpha Galaxies in the Epoch of Reionization \citep[LAGER,][]{Zheng2017} survey.

In this paper, we take full advantage of this serendipitous combination of telescope, instrument, and target and perform a LAE search around the quasar VIK~J2348--3054, with a bolometric luminosity and black hole mass of $4.3 \times 10^{46}$ erg s$^{-1}$ and $1.98 \times 10^{9}$ $M_{\odot}$ respectively, both of which are very close to the median values of all known quasars at this redshift (see Table 8 in \cite{Chiara2017b} and Table 1 in \cite{Chiara2023}. The remainder of the paper is organized as follows: Section 2 describes the observational setup adopted to perform the search. We present the results in Section 3, and in Section 4 we discuss the implications of these results and offer an explanation as to the current tension within the literature. We present our conclusions in Section 5. 

Throughout this paper we adopt a vanilla, flat $\Lambda$CDM cosmology with $\Omega_m =0.3$, and $H_0=70$ km s$^{-1}$ Mpc$^{-1}$. At $z=6.902$ the Universe is 0.753 Gyr old and the transverse scales are 24 arcseconds/cMpc and 189 arcseconds/pMpc. All magnitudes are in the AB system.

%--------------------------------------------------------------------

\section{Observations} \label{sec:data}
\subsection{DECam Observations of VIK J2348--3054}
\begin{figure}
    \centering
    \includegraphics[width = 0.5\textwidth]{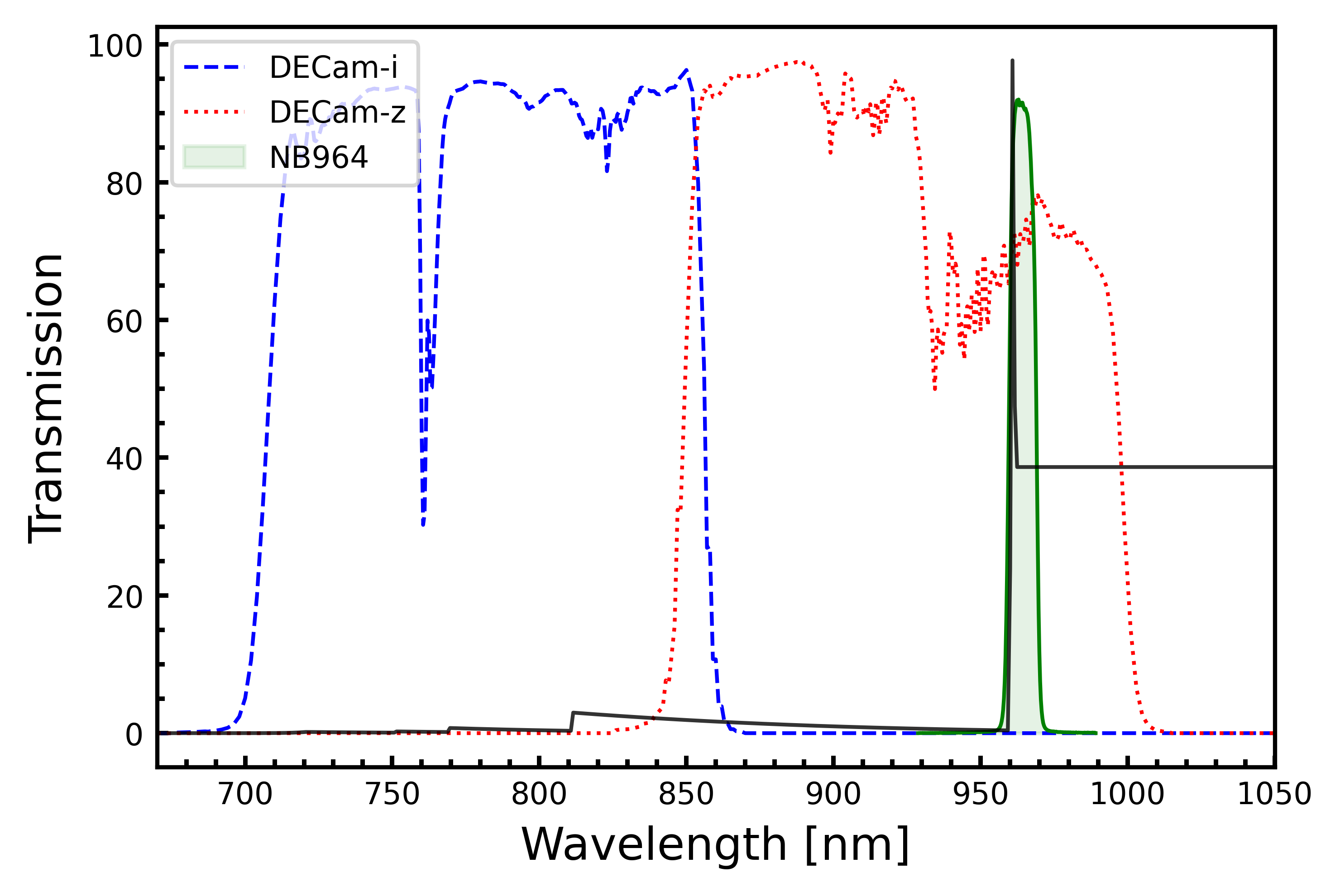}
    \caption{Transmission curves for the filters used in this work. The narrowband (NB964 filter) is represented by a solid green line and the two broadbands are represented as the blue and red dashed lines, for i and z, respectively. The LAE synthetic spectrum is shown as the solid black line.}
    \label{fig:curves}
\end{figure}
We observed the area around VIK J2348--3054 with DECam housed on the 4m Blanco telescope at the CTIO (PROPID: 2021B-0905). The observing run was split into two parts: the first took place over four half-nights from the UT 2021-08-30 to the UT 2021-09-03 where the moon did not rise during the observations, while the second occurred over two half-nights on UT 2021-10-20 and 2021-10-21 during full-moon. All observations were done under acceptable observing conditions and all frames were used. From the combined seven half-nights of observations we obtained integrations of 5.8, 8.6, and 14.2 hours in the i-band ($\lambda_{c} = 7840$ \AA; $\Delta \lambda$ = 1470 \AA), z-band ($\lambda_{c} = 9260$ \AA; $\Delta \lambda = 1520$ \AA), and NB964 narrowband ($\lambda_{c} = 9640 $ \AA; $\Delta \lambda$ = 94 \AA), respectively. The filter transmission curves are shown in Figure \ref{fig:curves}. 

This particular choice of filters allow us to search for LAEs
within a redshift range of 6.89 < z < 6.97, making it well suited
for a search around VIK J2348–3054 at z = 6.9018 \citep{Venemans2016}. Given that the Ly-$\alpha$ line tends to be redshifted
from the galaxy’s systemic velocity by $\sim$ 200 km s$^{-1}$ \citep{Verhamme2018}, we can detect galaxies with peculiar velocities ranging from $\sim$ -700 km s$^{-1}$ to $\sim$ 2250 km s$^{-1}$
from the quasar’s systemic velocity. Sources around high-z quasars and in protocluster environments can have peculiar velocities within 1000 $\pm$ km s$^{-1}$ \citep{Overzier2009}. This means that we should detect all sources associated with the protocluster environment with redshifted peculiar velocities, but may miss some sources with extreme blue shifted peculiar velocities.

DECam has 64 individual CCDs with numerous gaps between them. To account for this and obtain the best possible data quality, we implemented a dithering pattern during the observations, randomly shifting the pointing position within a 120 arcsecond box for each set of exposures. The pointing was also offset globally by 5 arcminutes in order to ensure that the quasar was not centered on a chip gap\footnote{The full observation script-generator is available at \url{https://github.com/TrystanScottLambert/DECam_Photometry/blob/main/scriptmaker.py}}.

Each band was stacked using the DECam community pipeline \citep{Valdes2014}. The DECam Community Pipeline provides the following operations: linearity correction, bias subtraction, dome, photometric, and dark sky flat fielding, bad pixel masking, masking of saturation effects and cosmic rays, an astrometric solution, removal of background patterns including stray light, pupil reflection, and fringing, and projection to a standard tangent plane. For multiple, dithered exposures a co-add is produced which includes transient masking. Three methods for masking cosmic rays and other transients were used in this work: the ``maskcosmics'' algorithm, from the Dark Energy Survey, using the traditional comparison to the PSF; an algorithm that identifies ``long'' cosmic rays based on the ratio of boundary to interior pixels from the IRAF segmentation and cataloging package; individual exposures are compared to a median by image differencing detection; and finally, statistical outlier rejection may be used when making the final stack.

After being stacked the images were astronomically aligned to the \textit{Gaia} DR2 catalog \citep{GAIA2018} and finally trimmed to cover the same area in all bands.

We masked several obvious CCD artifacts, which included stellar haloes and image ``ringing'', i.e., clusters of two or three unrealistically bright and/or negative pixels, by eye on a case by case basis. We also masked noisy edges by creating a polygon region within the borders of the combined images. The final usable field of view (FoV) is 2.87 deg$^2$. At the redshift of VIK J2348--3054 this area is equivalent to 1034~pMpc$^2$ 

\subsubsection{Photometric Calibration}
\begin{figure}
    \centering
    \includegraphics[width = 0.5\textwidth]{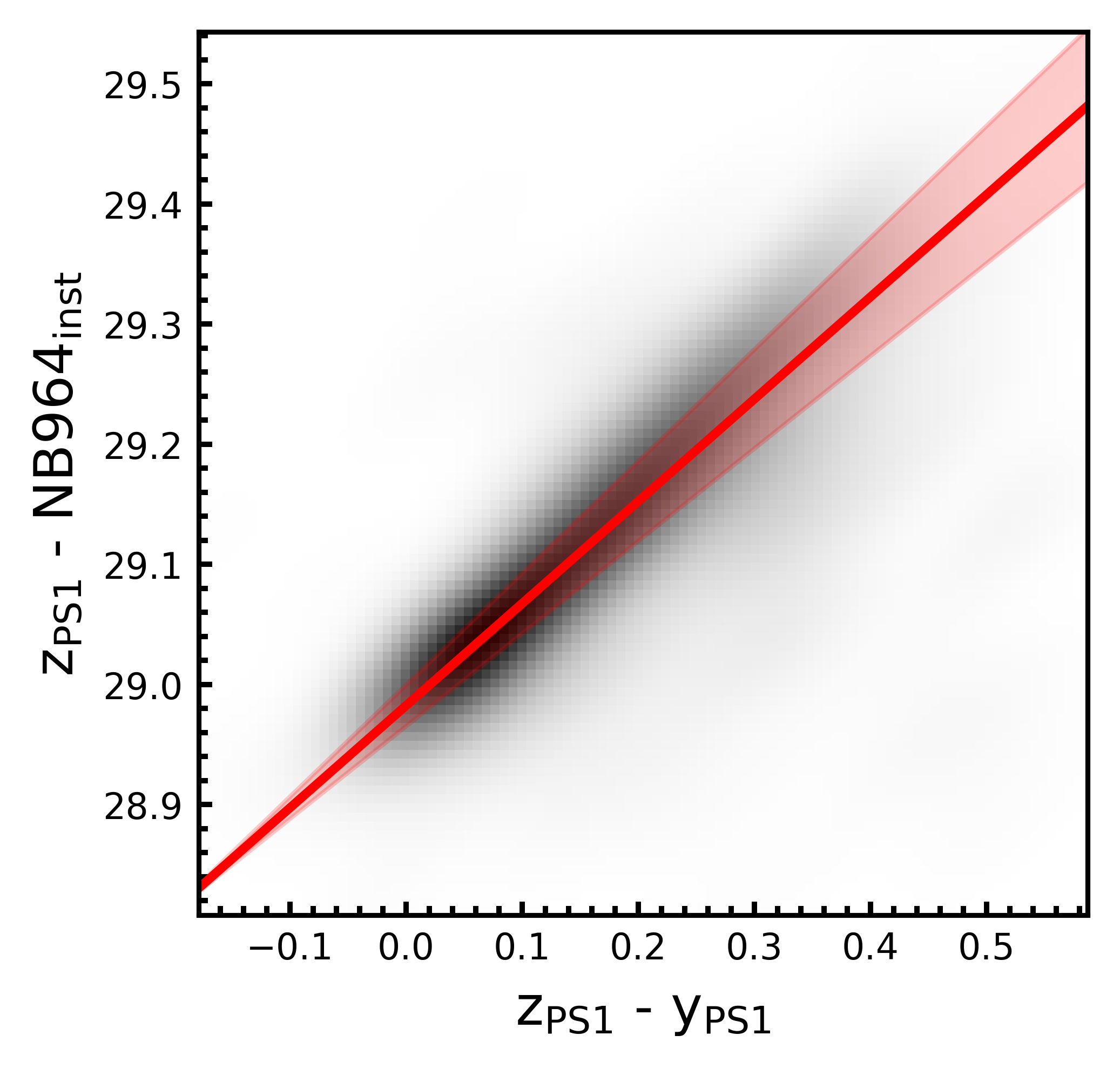}
    \caption{Color-color diagram of stars identified in the DECam image. Red line is a straight-line fit, indicating a linear relationship.}
    \label{fig:NB964}
\end{figure}

Previous DECam NB964 filter calibrations have often used Spectral Energy Distribution (SED) fitting of bright stars (either A, B or F0 stars) identified in the field \citep{Wold2019, Hu2019, Khostovan2020}. However, \cite{Wold2019} showed that calculating zero points assuming a linear relationship between the $z-NB964$ and $z-y$ colors, for bright, point like sources, is equivalent to SED fitting. We show this relationship for our data in Figure \ref{fig:NB964} by plotting the color-color diagram of sources which were identified as point sources by SExtractor \citep{sextractor}, and had z-band, y-band and i-band colors in the Panoramic Survey Telescope and Rapid Response System (Pan-STARRS1) first data release \citep{PANSTARS1}. We therefore photometrically calibrated the NB964 filter using the method by \cite{Wold2019, Wold2022}, assuming a linear relation between the PS1 and DECam magnitudes for point sources:
\begin{equation}
    z_{ps1} = \text{NB964}_{\rm inst} + \alpha \left(z_{ \rm ps1} - y_{ \rm ps1}\right) + \text{ZPT},
\end{equation}
where $z_{ps1}$ and $y_{ps1}$ are the $z$ and $y$ Pan-STARRS1 magnitudes respectively, NB964$_{inst}$ is the instrumental narrowband magnitude in DECam,  and $\alpha$ and ZPT are the slope and zero point which are needed to transform the magnitudes. The $\alpha$ and ZPT values are solved by fitting a best-fit straight line to 685 identified point sources, shown in Figure \ref{fig:NB964}. They where determined to be $0.9 \pm 0.01$ and $29.0 \pm 0.003$ respectively. The shaded area represents the $1\sigma$ value of the linear fit.
%The zero point value for 2" apertures was determined to be 29.14. 

The photometric transformations for the DECam broadband filters are available  via the Dark Energy Survey data management \citep{DES2021} who used (amongst other surveys) the Pan-STARRS1 survey \footnote{\url{https://des.ncsa.illinois.edu/releases/dr2/dr2-docs/dr2-transformations}}. Following their work, we photometrically calibrated the broadband filters using the following transformations:
\begin{align}
    z_{\rm DECam} &= z_{\rm ps1} -0.114 \left(r_{\rm ps1} - i_{\rm ps1}\right) - 0.010 \\
    i_{\rm DECam} &= i_{\rm ps1} -0.155 \left(r_{\rm ps1} - i_{\rm ps1}\right) + 0.015.
\end{align}

The offset between the instrumental DECam magnitudes and those estimated from the  Pan-STARRS1 magnitudes are used as the respective zero points.
%\begin{figure}
%    \centering
%    \includegraphics[width = 0.5\textwidth]{i_zpt.png}
%    \caption{i-band}
%    \label{fig:i_zpt}
%\end{figure}

%\begin{figure}
%    \centering
%    \includegraphics[width = 0.5\textwidth]{z_zpt.png}
%    \caption{z-band}
%    \label{fig:z_zpt}
%\end{figure}

%The zero points for the i-band and the z-band were determined to be 30.73, and 30.29 respectively.

We used SExtractor to create the source catalog for each DECam image. Because we are interested in LAEs, which must necessarily be detected in the narrowband, we use it as the detection image and then ran SExtractor in dual mode. The seeing values for the i, z, and narrow band were 1.17", 1.23", and 1.47" respectively, therefore we chose 2" diameter apertures. This choice of aperture size was also chosen during previous observations using the DECam instrument by \cite{Hu2019} who noted that this allows a large amount of flux to be observed whilst at the same time minimizing contamination from nearby sources. At the same time, this aperture choice is also larger than our worst seeing---1.47" for the narrowband.

The DECam pipeline generates an inverse variance image which takes into account pixel to pixel variations with respect to their uncertainties. We passed this weight-image as a parameter to SExtrator to estimate the photometric errors.

If a source was not detected in either broadband image, we adopted a 1$\sigma$ limiting magnitude and use this as a lower limit (27.14 for the i-band and 27.00 for the z-band). 

\subsubsection{Depth}
The limiting magnitudes were determined by fitting an exponential function to the relationship between detected sources' magnitudes and their signal-to-noise ratios (S/N), given by:
\begin{equation}
    S/N = \frac{2.5}{\ln\left(10\right)u\left(\text{mag}\right)},
\end{equation}
where $u\left(\text{mag}\right)$ is the photometric uncertainty, corresponding to the  \texttt{MAGERR$\_$APER} value from SExtractor.

We found that the limiting 5$\sigma$ magnitudes for the narrowband, i-band, and z-band were 24.65, 25.40, and 25.25 respectively, for 2" apertures.

\subsection{DECam Observations of the CDFS LAGER field}
We use the Lyman Alpha Galaxies in the Epoch of Reionization (LAGER) CDFS field observations from \cite{Hu2019}, for comparison since it does not contain a known quasar or any other feature that would make it unusual at the redshift of VIK J2348--3054. This field was observed with the same telescope, instrument, and filters that we used to search for LAEs around VIK J2348--3054. The CDFS images have several CCD artifacts as well as large areas of low S/N, mainly within the chip gaps due to insufficient dithering. We manually masked CCD artifacts and automatically masked the chip gap regions in each image using the corresponding weights image. The effective area available for source detection after masking was $\sim 1.56$ deg$^2$, or $\sim 570$~pMpc$^2$ at the redshift of VIK J2348--3054. 

We photometrically calibrate the CDFS images with the same method used for our DECam data.
%The zero points were determined to be 29.07, 31.33, and 31.43 for the narrowband, i-band, and z-band CDFS images respectively.

While the narrowband depths are similar between the observations around VIK J2348--3054 and CDFS, the broadbands in the latter are about 2 magnitudes deeper \citep[see ][]{Hu2019}. To create a comparison sample with the same photometric properties, it is necessary to degrade the CDFS photometry to mimic our depths. To do this we create a source catalog using SExtractor in the same way that we did for our data and then assign new errors to every detected source based on the fitted exponential functions from the depth calculations for our data (discussed Sect. 2.1.3) resulting in a degraded magnitude error, $\sigma_{\rm degraded}$. To replicate the expected scatter due to a lower depth image, we randomly assign a new source magnitude which is drawn from a normal distribution with a mean equal to the magnitude determined in the CDFS image and a standard deviation equal to $\sqrt{\sigma_{\rm degraded}^2 - \sigma_{\rm CDFS}^2}$, where $\sigma_{\rm CDFS}$ is the original magnitude error determined by SExtractor in the CDFS image. 

\subsection{Selection Criteria}

\begin{figure}
    \centering
    \includegraphics[width=0.5\textwidth]{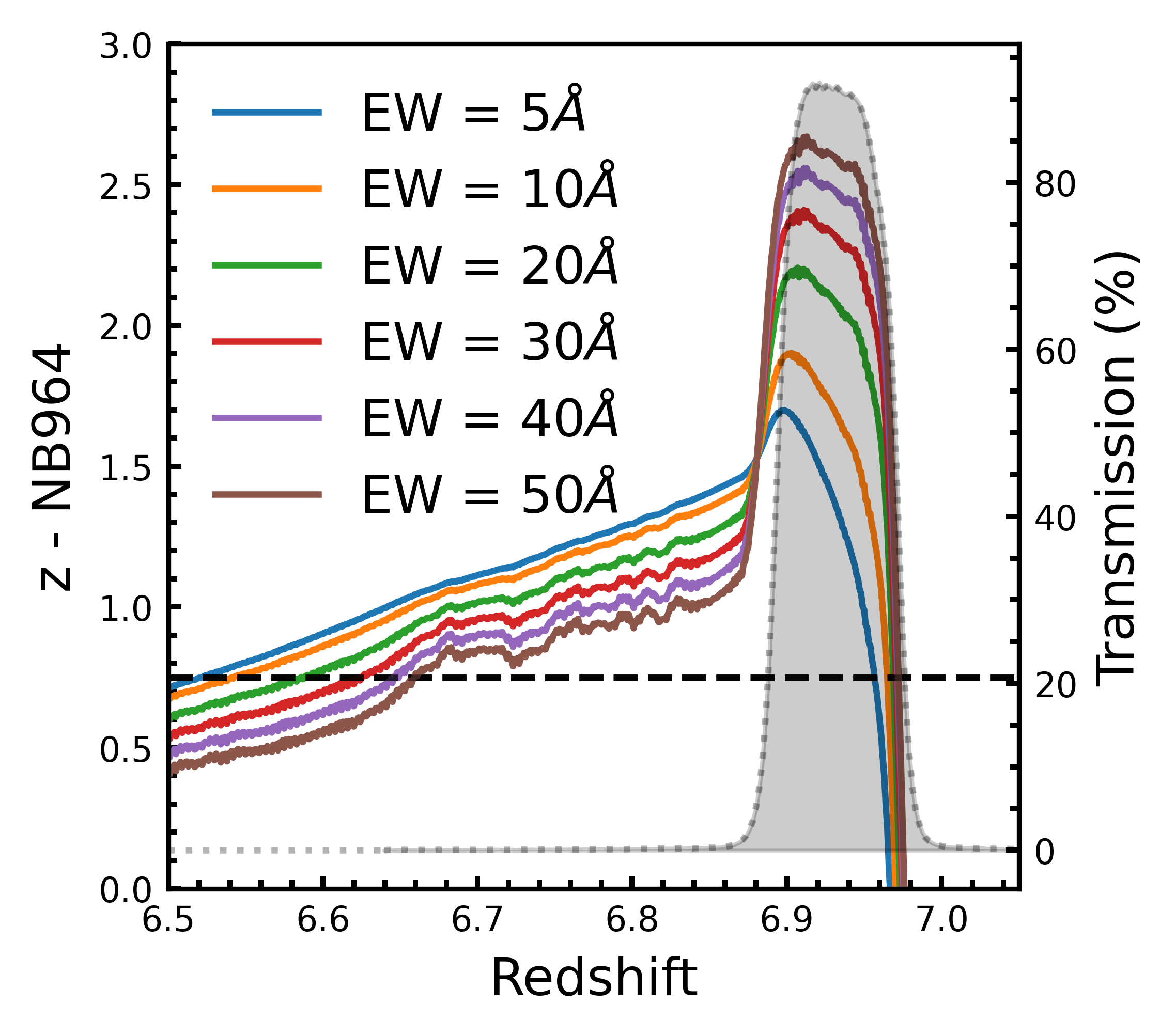}
    \caption{Predicted Z-NB964 color of a Ly-$\alpha$ emitter for redshifts ranging from 6.5 to 7 for varying equivalent widths. The dashed line indicates the Z-NB964 color selection criteria that we have adopted and the shaded region shows the NB964 filter profile.}
    \label{fig:hu2}
\end{figure}

A typical LAE has very strong Ly-$\alpha$ emission and a faint continuum \citep{Malhotra2002, Ouchi2020}. We plot a synthetic LAE spectrum at the redshift of VIK J2348--3054, along with the filter curves in Figure \ref{fig:curves}. The standard IGM extinction from \cite{Madau1995} was assumed. The line was made by assuming a Gaussian profile and adopting a 200 km s$^{-1}$ FWHM value and an equivalent width of 50 \AA. The luminosity is the L$^{*}$ and was derived using the $\alpha=-1.5$ luminosity function, reported in \cite{matthee2015}. As expected, the strong Ly-$\alpha$ emission is contained within the NB964 filter, with practically no flux in the i-band, while the z-band is dominated by the source continuum. Given this, we adopt a similar selection criteria to that used by \cite{Eduardo2013} and \cite{Chiara2017}, namely:

\begin{align}
    |z-NB| &> 2.5 \sqrt{\sigma_z^2 + \sigma_{NB}^2}, \\
    z-NB &> 0.78, \\
    i-z &> 1, \\
    \text{SNR}(i) &< 2.
\end{align}

Note that when written like this, the requirement (Eq. 5) deviates at faint magnitudes from a strict lower limit in the $f_{\rm NB}/f_z$ flux ratio and in practice ends up requiring a detection in the z-band. This makes the color selection significantly more robust against interlopers.

We also require that the Ly-$\alpha$ emission within the narrowband is bright with respect to the continuum (Eq. 6). Figure \ref{fig:hu2} shows the estimated $z$-NB964 color over a range of redshifts for LAEs with different equivalent widths. For all of them, the $z$-NB964 color is above 1.3. The criterion in Eq. (5) implies that a source with a color of 1.3 requires a corresponding error of $<0.52$, allowing us to capture LAEs within the photometric scatter without incurring in significant contamination (see section 4.1).

Finally, we require a significant identification of the continuum break short of Ly-$\alpha$ (Eq. 7). Figure \ref{fig:curves} shows clearly that the expected continuum emission on the blue side of the Ly-$\alpha$ line is very weak due to the IGM absorption in combination with the Lyman break, whilst the continuum on the red side should still be detected in the z-band. Our chosen criterion of i-z>1 enforces this and limits the numbers of interlopers (see section 4.1). We also require SNR($i$) $< 2$ (Eq. 8). 

\begin{figure}
    \centering
    \includegraphics[width=0.5\textwidth]{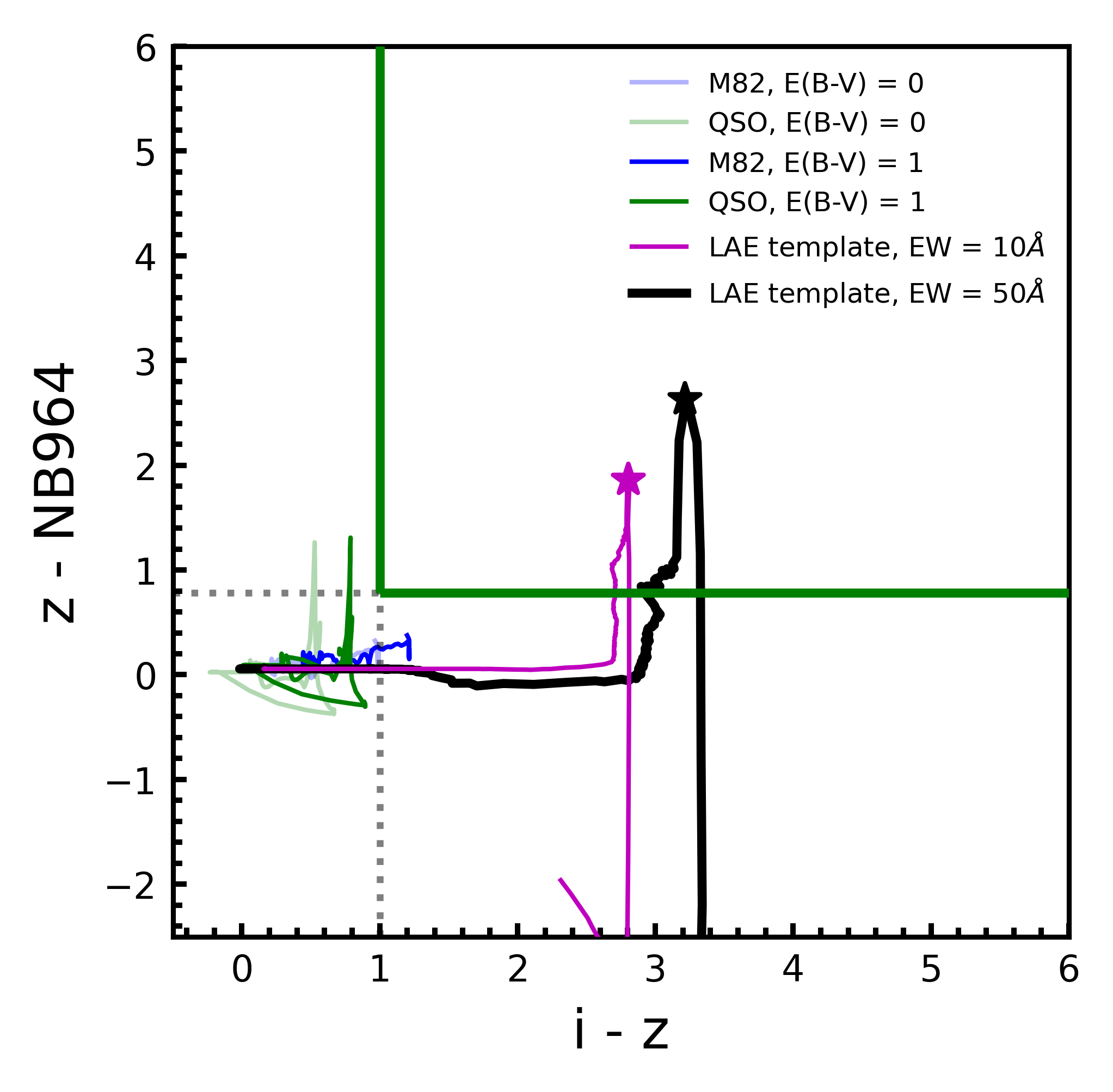}
    \caption{Color-color tracks of three model spectra: M82 (blue) and QSO2 (green) from \cite{Polletta2007}, the solid and transparent lines indicate an extinction E(B-V) = 1 and 0 respectively; the purple and black lines show the tracks of a model LAE spectrum with equivalent widths of 10\AA\ and 50\AA\ respectively. The stars represents the redshift of the quasar ($z=6.9$)}
    \label{fig:color-color track}
\end{figure}

To further justify our selection we take the synthetic LAE spectrum from Figure \ref{fig:curves}, and derive the expected colors using the python package \texttt{synphot}\footnote{Version 1.2.1} in the redshift range $5<z<7.2$. This color-color track is shown in Figure \ref{fig:color-color track} for equivalent widths of 10\AA\ and 50\AA. The Figure shows that the LAEs meet the $i-z$ criteria at the targeted redshift for all EWs of Lyman alpha considered in Figure \ref{fig:hu2}. At the same time we tested two types of interlopers using the spectral energy distributions provided in \cite{Polletta2007}: M82, and QSO2. These represent a star forming galaxy and a type-2 quasar respectively. We plot their color-color tracks from $0<z< 1.3$ in green and blue and include no dust extinction and E(B-V) = 1. Neither of these two low-z interlopers enter our color selection (indicated as the top right green box in Figure \ref{fig:color-color track}).  This justifies our color selections, demonstrating that these choices are optimal to select Ly-$\alpha$ emitters around the quasar redshift. 

\begin{figure}
    \centering
    \includegraphics[width=0.5\textwidth]{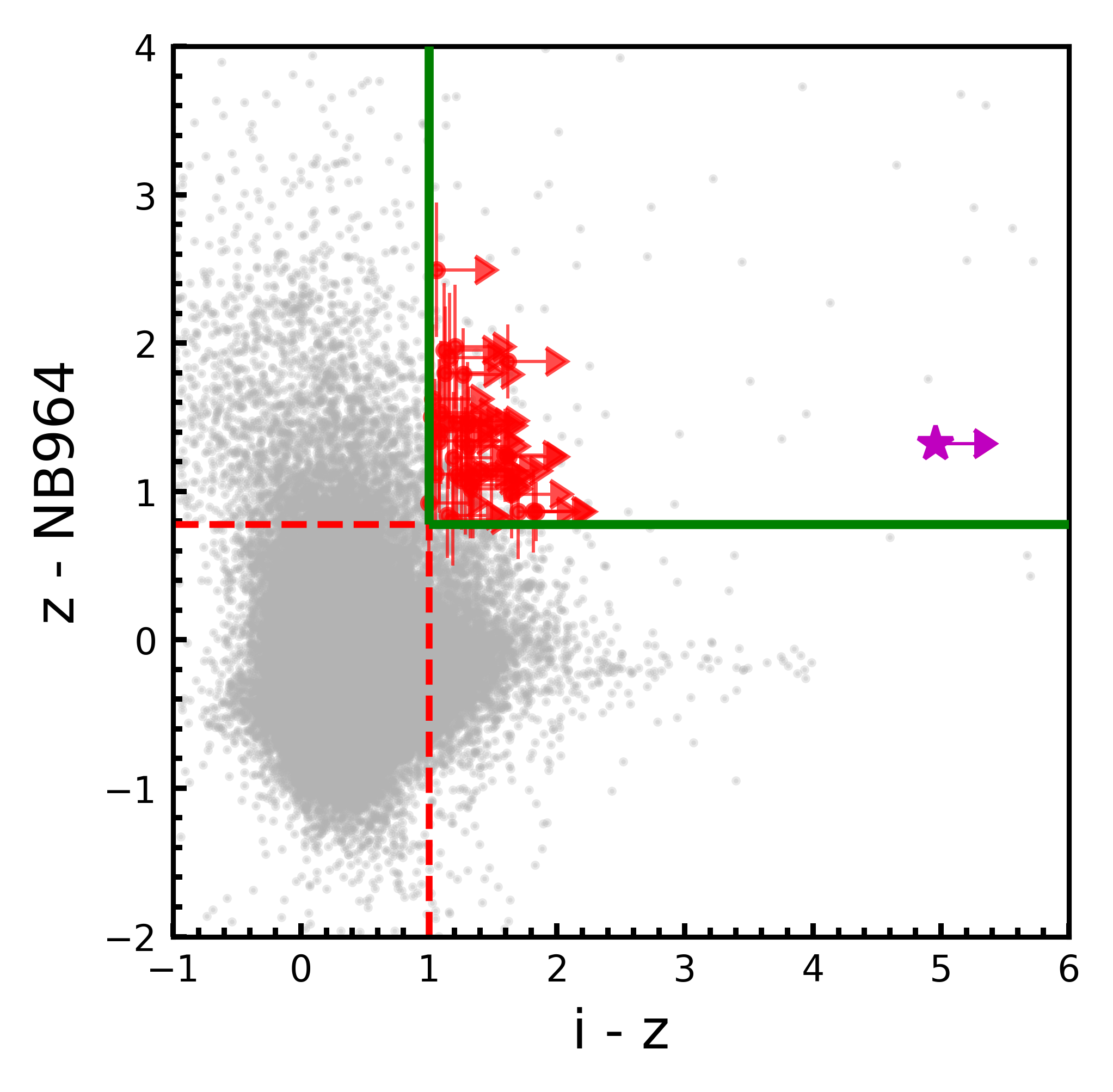}
    \caption{Color-color diagram. Red points show the LAE candidates in the field of VIK J2348--3054. Grey points show all the sources that were identified with SExtractor. The purple star is the position of the quasar. Gray points in the selection region are not Ly-$\alpha$ candidates because they failed either Equations (5) or (8).}
    \label{fig:color_color}
\end{figure}

The color-color plot of all the detected sources in our field is shown in Figure \ref{fig:color_color}. The grey points are the detected sources, the red dashed lines indicated the color selection criteria, and the red points highlight the LAE candidates which meet the selection criteria. The green box highlights the color selection for LAEs.

\section{Results} 
\subsection{LAE Candidates}
We find 39 sources sources in the VIK J2348--3054 field which meet our selection criteria: 38 LAE candidates as well as the quasar itself. The 20 pix $\times$ 20 pix (10.8" $\times$ 10.8") postage stamp cut outs of all the candidates are presented in Figure \ref{fig: LAE-candidates} in the appendix. We also indicate the magnitudes in the respective bands as well as their S/N. While all candidates formally meet our selection criteria (see Sect. 2.3), we do note that three candidates (LAE-25, LAE-30, and LAE-38) seem to have obvious i-band detections despite SExtractor assigning them very low S/N values. By contrast, only two candidates were identified in the CDFS images after the photometry degradation (see Sect. 2.3). 

Our candidate properties are presented in Table \ref{tbl:catalogue} in the appendix, including their positions, separation from  VIK J2348--3054,  i-band, z-band, and NB964 magnitudes, Ly-$\alpha$ luminosities, and star formation rates  (see sections 3.2 and 3.3). VIK2348--3054 also met the selection criteria and is therefore also presented in Table 1. The z magnitude determined for the quasar is nearly 0.8 mag fainter than the one reported in \cite{Venemans2013}, upon discovery, who measured a Gunn z filter magnitude of $22.96\pm0.18$ on the 3.8m New Technology Telescope (NTT). The slight difference between the two is easily explainable by the difference in filter sets between DECam and NTT; the z-band in DECam is much redder than in EFOCS2 (because of the lower efficiency of the CCD at long wavelengths). This is also notable in Table 3 of \cite{Chiara2017b} where the PS1 z-band magnitudes differ from the DECam z-band photometry by substantial amounts due to the filter curve. Additionally, our DECam z value is also expected given the relationship between previous quasars' DECam z magnitudes and their J band magnitudes \citep[see Table in 3 in][]{Chiara2017} which have comparable redshifts and luminosities to VIKJ2348-3054.

\begin{figure*}
\sidecaption
    \centering
    \includegraphics[width=12cm]{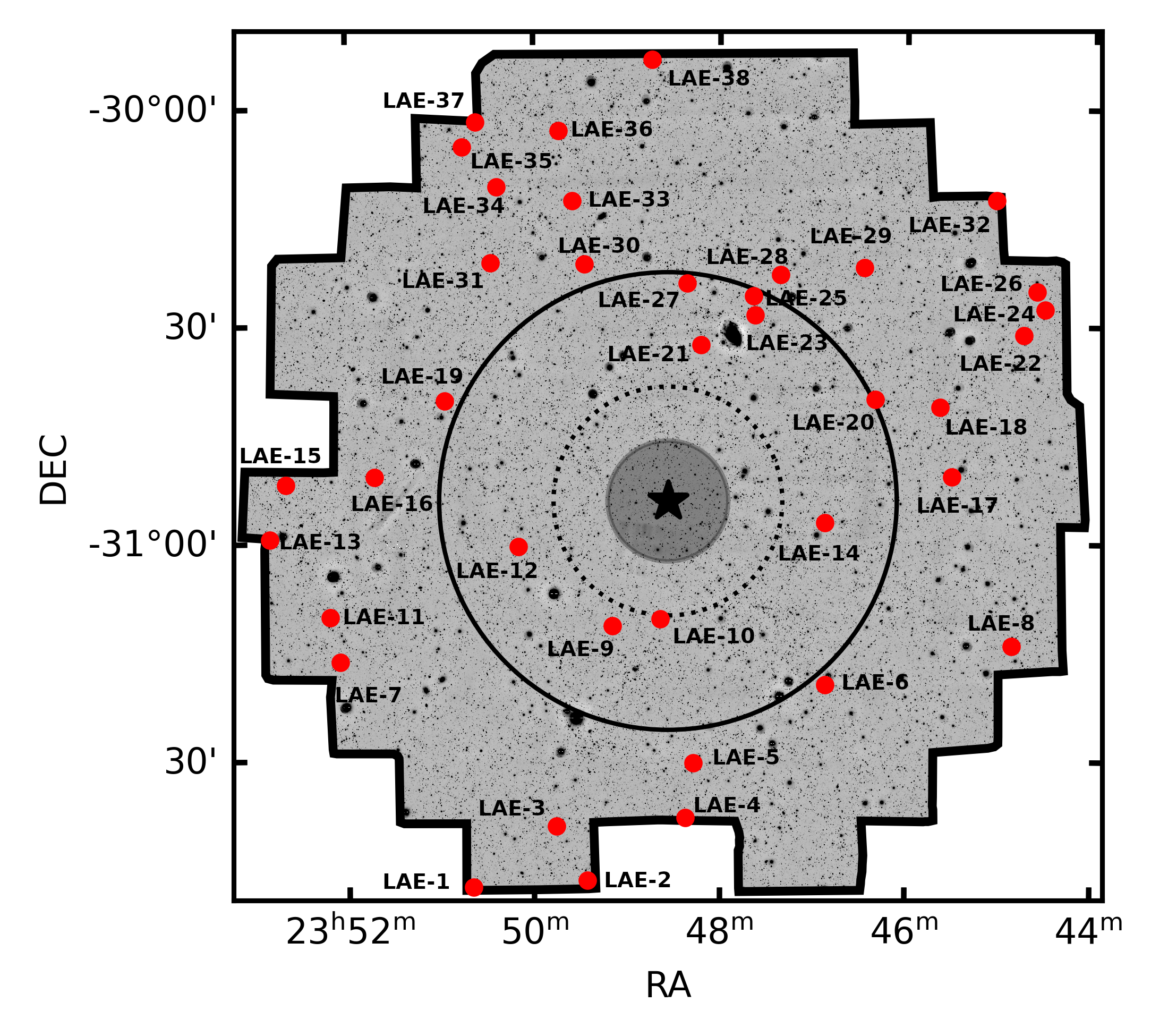}
    \caption{LAE on-sky distribution around the quasar. Background image is the stacked narrowband DECam image around the quasar. The black outline is the area used to identify LAE and is slightly smaller than the full DECam FOV. The star at the image center is the quasar location. The shaded region shows the proximity zone determined by \cite{Chiara2017b}. The inner and outer rings highlight 5~pMpc and 10~pMpc from the quasar respectively, with the later being the maximum scale at which objects are expected to collapse and form clusters at $z=0$ \citep{Overzier2009}.}
    \label{fig:on_sky_distribution_us}
\end{figure*}

The on-sky distribution of the LAE candidates around the quasar can be seen in Figure \ref{fig:on_sky_distribution_us}.

%\begin{figure}
%    \centering
%    \includegraphics[width=0.5\textwidth]{candidate_cdfs1.png}
    %\includegraphics[width=0.5\textwidth]{candidate_cdfs2.png}
    %\caption{LAE candidate 20 pix $\times$ 20 pix (10.8" $\times$ 10.8") cutouts, identified in the degraded CDFS data. The values in the top of each postage stamp indicate the AB magnitudes in the i, NB964, and z bands whilst the bottom show the S/N values in those respective bands.}
    %\label{fig: CDFS-candidates}
%\end{figure}

\subsection{Estimation of Ly-$\alpha$ Luminosities}

We follow the same method used by \cite{Hu2019} to determine the Ly-$\alpha$ luminosities of our candidates. Specifically, we use the following equation to estimate the flux density from the Ly-$\alpha$ emission line ($f_{\lambda, \text{line}}$),

\begin{align}
    k_{x} \bar{f_{\nu}} &= \int \left(f_{\lambda, \text{line}} + f_{\lambda, \text{cont}}\right) T_{x}(\lambda)d\lambda \\
    &= \int f_{\lambda, \text{line}}T_{x}(\lambda)d\lambda + \int f_{\lambda, \text{cont}} T_{x}(\lambda) d\lambda,
\label{eq: hu4}
\end{align}
where $T_{x}(\lambda)$ is the transmission function of filter $x$ (in our case $x$ is either NB964 or the z-band), $\bar{f_{\nu}}$ is the observed flux density, and $k_{x}$ is a constant defined as:
\begin{equation}
    k_{x} \equiv c\frac{\int T_{x}(\lambda) d\lambda}{\bar{\lambda_{x}}^2},
\end{equation}
where $\bar{\lambda_{x}}$ is the central wavelength of filter $x$ and $c$ is the speed of light. For simplicity, we assume that the Ly-$\alpha$ emission line ($f_{\lambda, \text{line}}$) is a $\delta$-function, centred on the NB964 filter, and the continuum flux ($f_{\lambda, \text{cont}}$) is a power law spectrum with a slope of -2 and attenuated by the IGM using the model from \cite{Madau1995}. Making these assumptions allows us to express Equation (\ref{eq: hu4}) as:
\begin{equation}
    k_{x}\bar{f_{\nu}} = f_{\alpha} T_{x}(\lambda_{\alpha}) + C \int \lambda^{-2} e^{-\tau} T_{x}(\lambda) d\lambda,
    \label{eq: generalized_form}
\end{equation}
where $f_{\alpha}$ is the Ly-$\alpha$ flux, $C$ is a constant, and $\tau$ is the average IGM absorption optical depth at the redshift of the quasar. When we substitute the narrowband and broadband values into Equation (\ref{eq: generalized_form}) we get two equations:
\begin{align}
    \label{eq:sim1}
    k_{NB} \bar{f}_{\nu, NB} &= f_{\alpha}T_{NB}(\lambda_{\alpha}) + C \int \lambda^{-2} e^{-\tau} T_{NB}(\lambda) d\lambda \\
    \label{eq: sim2}
    k_{BB} \bar{f}_{\nu, BB} &= f_{\alpha}T_{BB}(\lambda_{\alpha}) + C \int \lambda^{-2} e^{-\tau} T_{BB}(\lambda) d\lambda,
\end{align}
where $\lambda_{\alpha}$ is the Ly-$\alpha$ emission wavelength. We can solve Equations \ref{eq:sim1} and \ref{eq: sim2} simultaneously for $C$ and $f_{\alpha}$, using the latter to determine the Ly-$\alpha$ luminosity ($L_{\text{Ly}\alpha}$).

\subsection{Star formation rates}
We estimate the star formation rates (SFRs) by following the same procedure used by \cite{Chiara2017}. Using the derived Ly-$\alpha$ luminosities ($L_{\text{Ly}\alpha}$), we can determine H$\alpha$ luminosities ($L_{\text{H}\alpha}$) by assuming case-B recombination and exploiting the relationship presented in \cite{Osterbrock1989}: $L_{\text{Ly}\alpha} = 8.7 \times L_{\text{H} \alpha}$. SFRs can be calculated using the relationship between SFR and $L_{\text{H}\alpha}$ presented in \cite{Kennicutt2012}:

\begin{equation}
    \log\left(\frac{\text{SFR}_{L_{\text{Ly}\alpha}}}{M_{\odot} \text{ yr}^{-1}}\right) = \log\left(\frac{L_{\text{H}\alpha}}{\text{erg} \text{ s}^{-1}}\right) - 41.27.
\end{equation}

However, it is worth noting that these star formation rate estimates are highly uncertain given that we do not know the intrinsic Ly-$\alpha$ luminosity due to star formation and the observed Ly-$\alpha$ luminosity is most likely a fraction of the intrinsic total.

\section{Discussion} \label{sec:discussion}
\subsection{Overdensity of LAEs}
Considering the number of candidates (section 3.1) and usable area of each field (section 2), we find the surface density of LAEs around VIK J2348--3054 to be $13.2\pm2.2$ deg$^{-2}$, whilst in CDFS we only find a surface density of $1.3 \pm 0.9$ deg$^{-2}$ to the same depth (see Sect. 2.3 for details). If we define $\delta=1$ as a density consistent with a field density, then this quasar is sitting within an overdense region with an overdensity factor of $\delta = 10^{+5.3}_{-5.9}$, where the uncertainty corresponds to the 68$\%$ confidence interval. We note that given this highly skewed distribution, the probability of the field not being overdense (i.e, the probability that $\delta \leq 1$) is 3$\times10^{-6}$. To further explore the overdensity factor and make sure the result is not driven by the CDFS density, we constructed a Ly-$\alpha$ luminosity function (see the values used in Table \ref{tbl: luminosity}) and compared it to the combined CDFS and COSMOS fields from \cite{Hu2019} and the combined LAGER 4-field luminosity function from \cite{Wold2022} in Figure \ref{fig:luminosity_function}. We define the luminosity function as:

\begin{equation}
\Phi\left(L\right) dL = \sum\limits_{L_i \in [L-\Delta L/2, L + \Delta L/2]} \frac{1}{V_{\rm eff} f_{\rm comp}}dL,
\end{equation}
where $V_{\rm eff}$ is the effective volume of our survey--$1.8\times10^{6}$~cMpc$^{3}$--determined by taking into account the survey area and the redshift range probed, and $\Delta L$ is the bin-width = 0.1 $\log L$. We also include a detection completeness factor---$f_{\rm comp}$---as described in Hu et al. (2019) and Wold et al. (2022). This factor is calculated by inserting, and then recovering, false sources in the narrowband. The $f_{\rm comp}$ factor is shown in Figure 8 as a function of magnitude. Specifically, we use the best-fit error function shown in the Figure to estimate the luminosity function. To further constrain which equivalent widths we are sensitive to we also investigated what range of equivalent widhts would pass our selection criteria for a given Ly-$\alpha$ luminosity. In particular: we assumed zero flux in the i-band; for each bin in Figure \ref{fig:luminosity_function} we calculated the percentage of synthetic LAEs that would pass our selection criteria, with equivalent widths ranging from 1 - 200 $\AA$; we then divide the luminosity function by this completeness correction factor, which results in the blue points in Figure \ref{fig:luminosity_function}. Our limit-assumption correction suggests an even larger overdensity. However, the uncommon shape of the LF observed after correction suggests that there is likely a more complex distribution of EWs, and the correct LF sits in between the black and blue dots.

While the overdensity is not clear in the two faintest bins, this may be due to incompleteness from our selection function. Further observations, both deeper and in other bands (as has been done for the LAGER survey, e.g., Hu et al. 2019) are likely required to determine the source of this. To estimate $\delta$, we fit the amplitude of the luminosity functions from \cite{Hu2019} and \cite{Wold2022} to our data, without considering the two faintest bins. We find that the candidates around VIK J2348--3054 have a higher space density. The luminosity functions from \cite{Hu2019} and \cite{Wold2022} are in tight agreement (as is evident in Figure \ref{fig:luminosity_function}), and both imply an overdensity of $\delta = 4  \pm 0.4$. These luminosity functions were corrected for selection incompleteness whereas our data has not been corrected in the same way. This has the consequence of the measured $\delta = 4 \pm 0.4$ value representing a lower limit, which again, indicates that the quasar lives in an overdense environment.

Interestingly, we are only able to find this overdensity due to the large size of our field, as there appears to be a distinct lack of sources in the vicinity of the quasar (see Figure \ref{fig:on_sky_distribution_us}). Figure \ref{fig:surface density} shows the LAE candidates' radial distribution from the quasar. The blue shaded region represents the average CDFS surface density of $1.3 \pm 0.9$ deg$^2$, with an uncertainty determined by considering the Poisson single-sided,  $1 \sigma$ upper limit for $n=2$ \citep{SmallNumberStatistics}. Radial bins with no detections have been given an upper limit of 1.846 objects which is again the $1\sigma$ upper limit for a single-sided Poisson distribution. Figure \ref{fig:surface density} also shows the maximum radius that objects are expected to collapse into a cluster at $z=0$, 75 cMpc (9.5 pMpc) expected at this redshift \citep{Overzier2009} and is also shown as the outer ring in Figure \ref{fig:on_sky_distribution_us}. Figure \ref{fig:surface density} shows that the field is overdense throughout all the area covered by our images, except within the inner ~5pMpc from the quasar. We discuss this further in the next section.

\begin{figure}
    \centering
    \includegraphics[width=0.5\textwidth]{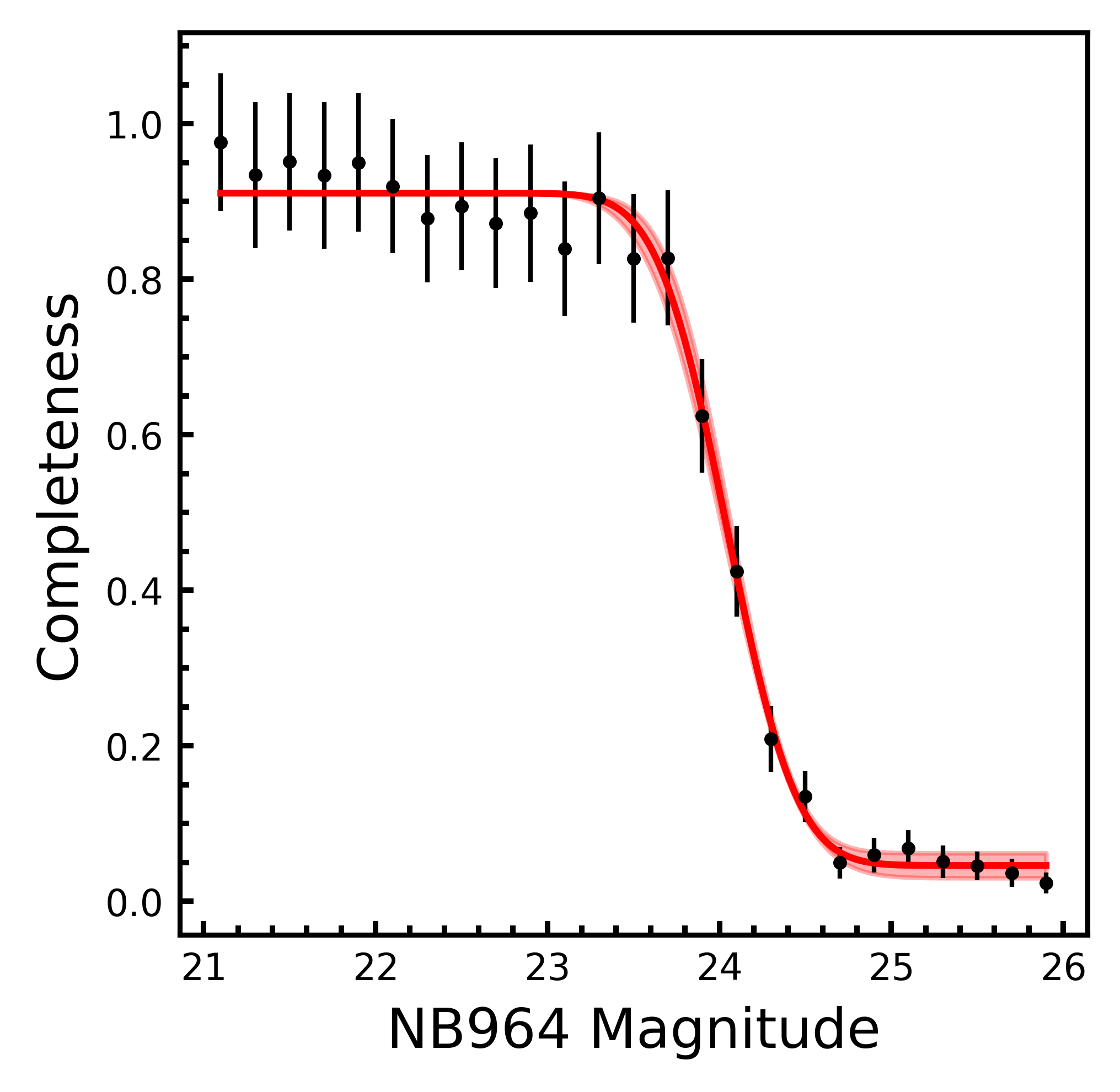}
    \caption{Narrowband completeness function. The red line shows the best-fit error function used in the luminosity function calculation (Eqn. [16]).}
    \label{fig: completeness}
\end{figure}

\begin{figure}
    \centering
    \includegraphics[width=0.5\textwidth]{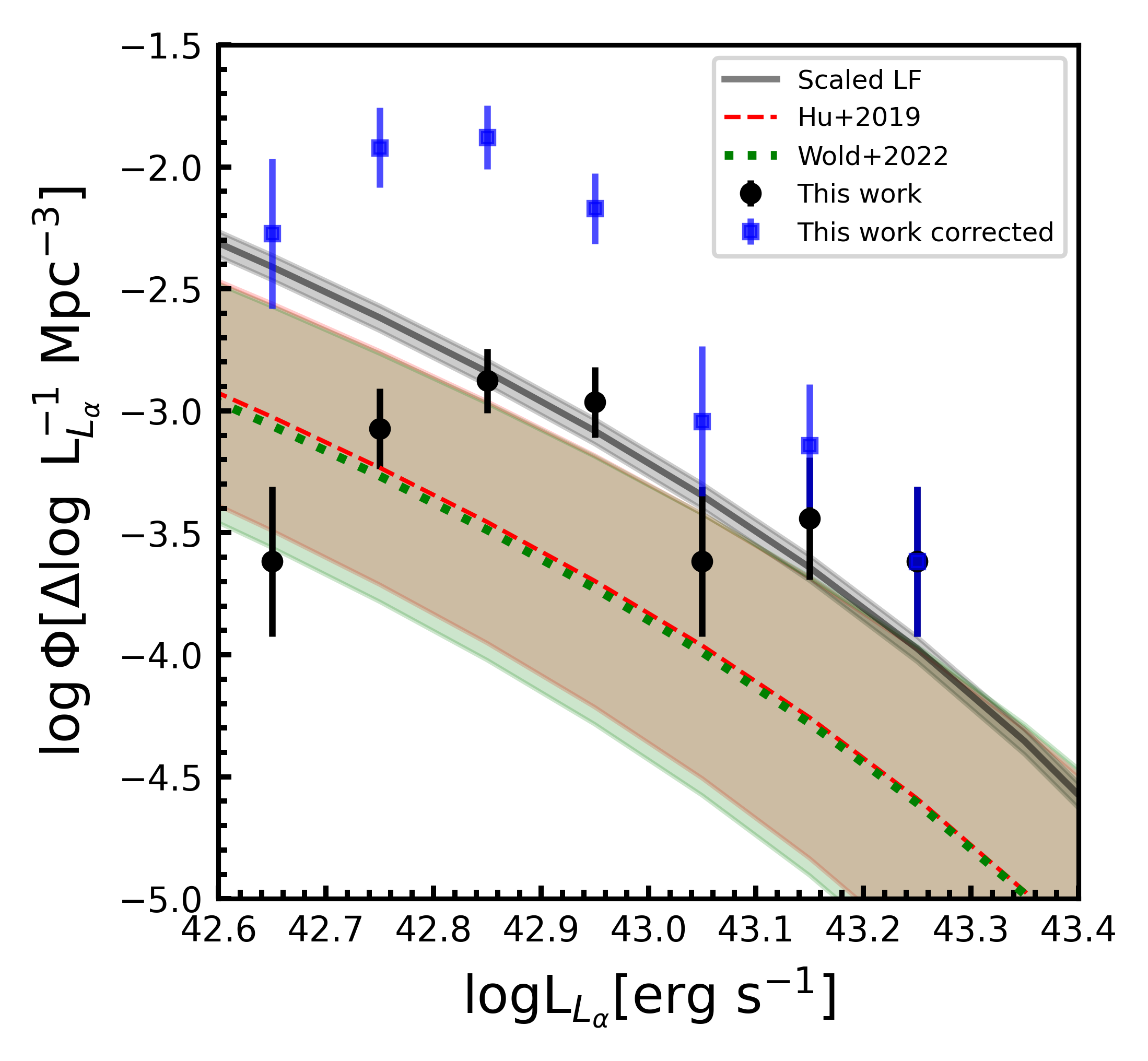}
    \caption{The red, dashed,line shows the selection incompleteness uncorrected luminosity function determined by \cite{Hu2019} using the combined COSMOS and CDFS fields. The green, dotted, line shows the combined luminosity function from \cite{Wold2022}. The black points are the values determined from the LAEs observations in our field. The black line is the \cite{Hu2019} luminosity function scaled to our data. Blue points are the limit-assumption completeness corrected luminosity function as described in section 4.1.}
    \label{fig:luminosity_function}
\end{figure}

\begin{table}[]
\caption{Counts, space density, and space density uncertainty for 0.1 luminosity bins.}
\label{tbl: luminosity}
\begin{tabular}{cccc}
\hline \hline \\
Luminosity Range                               & Counts & $\log \phi$                & $u(\log \phi)$          \\
$\log L_{L_{\alpha}}$ [erg s$^{-1}]$ &        & $\Delta \log L_{L_{\alpha}}$ Mpc$^{-3}$ & $\Delta \log L_{L_{\alpha}}$ Mpc$^{-3}$ \\ \hline
42.6 -- 42.7                                   & 2      & -3.62                                 & 0.31                                  \\
42.7 -- 42.8                                   & 7      & -3.07                                 & 0.16                                  \\
42.8 -- 42.9                                   & 11     & -2.88                                 & 0.13                                  \\
42.9 -- 43.0                                   & 9      & -2.96                                 & 0.14                                  \\
43.0 -- 43.1                                   & 2      & -3.61                                 & 0.31                                  \\
43.1 -- 43.2                                   & 3      & -3.44                                 & 0.25                                  \\
43.2 -- 43.3                                   & 2      & -3.62                                 & 0.31                                  \\
43.3 -- 43.4                                   & 0      & -                                     & -                                     \\
43.4 -- 43.5                                   & 2      & -3.62                                 & 0.31                                  \\
43.5 -- 43.6                                   & 0      & -                                     & -                                     \\
43.6 -- 43.7                                   & 0      & -                                     & -        \\ \hline                            
\end{tabular}
\end{table}

\begin{figure}
    \centering
    \includegraphics[width=0.56\textwidth]{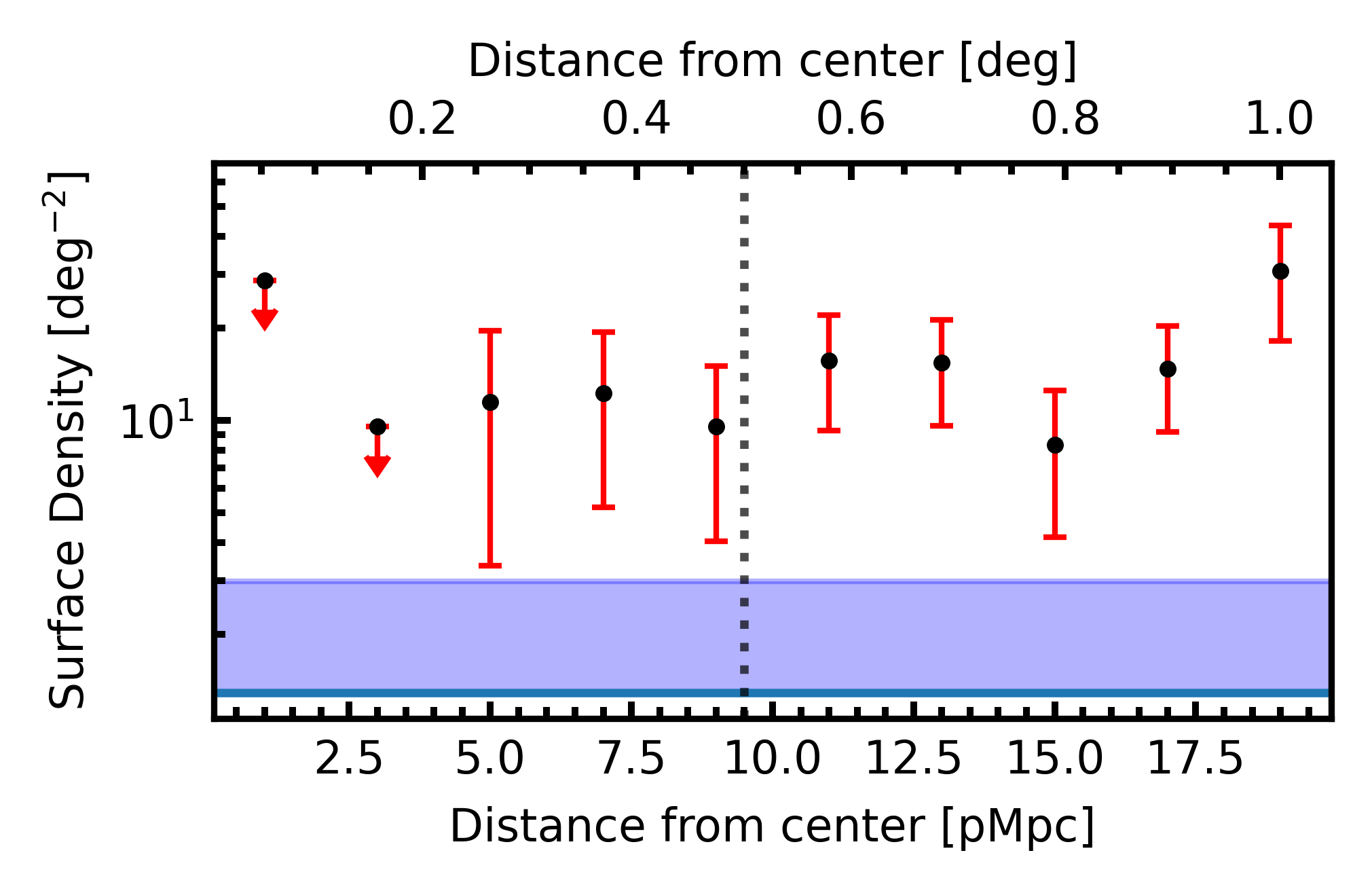}
    \caption{Surface density profile as a function of distance from the central quasar. The data were binned using 2 pMpc bins. Red arrows indicate the 1 $\sigma$ upper limits based on non-detections and assuming a Poisson uncertainty. The blue shaded region is the average upper limit for the degraded CDFS data. The dashed line is the maximum theoretical collapse radius found by \cite{Overzier2009}.}
    \label{fig:surface density}
\end{figure}

\subsection{Central suppression of Ly-$\alpha$ emitters}
Figure \ref{fig:on_sky_distribution_us} shows the on-sky distribution of LAE candidates. The inner shaded region represents the quasar proximity zone---the theoretical region around a quasar where its UV radiation would have ionized a pure hydrogen ISM \citep{Fan2006}. \cite{Chiara2017b} determined the radius of the proximity zone of VIK J2348--3054 to be 2.64~pMpc. The Figure also shows ring with the largest radius that \cite{Overzier2009} found to contain overdensities. It is worth noting that there is a small observational artifact in the form of striations, to the south of the quasar, which overlaps the proximity zone. This is caused by a relatively strong main reflection pattern in DECam, possibly by moonlight. However, the area affected is very small and this effect had no impact on LAE candidate selection.

What is striking about the on-sky distribution is the discernible lack of candidates towards the center of the image, around the quasar position (also see Figure \ref{fig:surface density}). In fact, the nearest candidate to the quasar is $\sim$ 0.27$^{\circ}$ away, i.e., 5.15~pMpc. This distance is indicated by the dashed ring in Figure \ref{fig:on_sky_distribution_us}. If we take the area outside of the maximum radius at which objects are expected to collapse into clusters at $z=0$, i.e, the region beyond the outer ring in Figure \ref{fig:on_sky_distribution_us}, where we do not expect the quasar to have much, if any, influence, and use it to estimate the average field density, then the Poisson probability of finding the inner region void of LAEs is 1.2$\%$. However, when coupled with the fact that this underdense region is centered on the already rare luminous high-z quasar, we can conclude it is highly unlikely that this is a chance occurrence, but instead shows a real dearth of LAEs. Furthermore, we would naively expect the quasar to be close to the center of overdensity, where the environment should be more concentrated than in the outskirts, making this underdensity even more striking. In the following sections we discuss potential explanations for this observed lack of LAEs in the vicinity of VIK 2348-3054. 

%We also investigate the on-sky distribution more quantitatively by calculating the two-point angular correlation function, defined by \cite{twopointcorrelation} as:
%\begin{equation}
%    \omega(\theta) = \frac{\text{DD}(\theta) - \text{2DR}(\theta) + \text{RR}(\theta)}{\text{RR}(\theta)},
%\end{equation}

%with uncertainties
%\begin{equation}
%    \sigma_{\omega} (\theta) = \frac{1 + \omega(\theta)}{\sqrt{\text{DD}(\theta)}}.
%\end{equation}

%DD$(\theta)$, DR$(\theta)$, and RR$(\theta)$ are the number of galaxy-galaxy, galaxy-random, and random-random pairs, at a given angular separation $(\theta)$. We show the angular two point correlation function for our data in Figure \ref{fig:actpf}. It is relatively flat except for a noticeable dip and then peak at $\sim$ 0.3$^{\circ}$, consistent with the distance between nearest LAE candidate (LAE-10) and the quasar. This suggests that the deviation from a uniform angular correlation is due to the lack of LAEs at the image center.

%\begin{figure}
%    \centering
%    \includegraphics[width = 0.5\textwidth]{angular_tpcf.png}
%    \caption{Angular two-point correlation function of LAE candidates observed in our field.}
%    \label{fig:actpf}
%\end{figure}

\subsubsection{Star formation suppression due to negative feedback}
A popular mechanism previously suggested to explain a lack of LAEs around some quasars has been negative feedback from the quasar itself \citep{Eduardo2013, Morselli2014, Chiara2017, Ota2018}. In this scenario, the strong ionizing radiation from the quasar photoevaporates neutral hydrogen in its immediate vicinity, prohibiting gas condensation, and thus inhibiting star formation in galaxies where it would otherwise be able to occur \citep{Kim2009, Overzier2009}. Our results are consistent with the expected scale proposed by a variety of studies \citep[e.g., ][]{Chen2020, Zhou2023}, all suggesting that this effect should be within $\sim 5$~pMpc. We further quantify this by calculating the isotropic UV intensity in a sphere around the quasar, following \cite{Kashikawa2007}, and determine the local flux density at different distances from the quasar via:

\begin{equation}
\label{eq: flux cont}
    F^{Q}_{\nu} = \frac{L(\nu_L)}{4\pi r^2},
\end{equation}
 where $r$ is the distance from the quasar and $L(\nu_L)$ is the quasar luminosity at the Lyman limit (912\AA). The latter can be estimated as in \cite{Chiara2017b} by using the extinction-corrected rest-frame 1450 \AA\ AB magnitude:

\begin{equation}
\label{eq: luminosity cont}
    L(\nu_L) = 4\pi D_{L}^2 f_{\nu}^0 10^{-0.4 m_{1450}} \left(\frac{912}{1450}\right)^{-\beta}
\end{equation}
 where $D_L$ is the luminosity distance to the quasar, $m_{1450}$ is the extinction-corrected rest-frame 1450 \AA\ AB magnitude, $f_\nu^0$ is 3631 Jy, and $\beta$ is the continuum slope \citep{Fan2001}.

We use the value of $m_{1450}=21.17$ mag estimated by \cite{Chiara2017b} for VIK~2348--3054 by extrapolating from the observed J-band magnitude. We also adopt $\beta=-0.99$ following \cite{Kashikawa2007}. Using Equations (\ref{eq: flux cont}) and (\ref{eq: luminosity cont}) we determine the local flux density at 2 pMpc and 5 pMpc to be:
\begin{equation}
    F_{\nu}^{Q} (\nu_L; 2 \text{pMpc}) = 9.0 \times 10^{-20} \text{erg cm$^{-2}$ s$^{-1}$ Hz$^{-1}$}
\end{equation}
and
\begin{equation}
    F_{\nu}^{Q} (\nu_L; 5 \text{pMpc}) = 1.4 \times 10^{-20} \text{erg cm$^{-2}$ s$^{-1}$ Hz$^{-1}$}.
\end{equation}
 The isotropic UV intensity at the Lyman limit is then $J_{21}\sim 7$ and $J_{21} \sim 1$ at 2~pMpc and 5~pMpc respectively, where $J_{21} = J/10^{-21}$ erg cm$^{-2}$ s$^{-1}$ Hz$^{-1}$ sr$^{-1}$ and $J$ is the UV intensity at the Lyman limit, determined by dividing the local flux density by $4\pi$ sr. 
 
 Several simulations suggest that the $J_{21} > 1$ is sufficient to completely inhibit star formations in haloes of mass below  $10^{9} M_{\odot}$ \citep{ThoulWeinberg1996,Kashikawa2007, Chen2020}. Given this, and the values we determined for $J_{21}$, it is physically possible these LAEs have been suppressed within the quasar proximity zone---2.64~pMpc \citep{Chiara2017b}, shown by the shaded region in Figure \ref{fig:on_sky_distribution_us}---up to where we find the first LAE at 5~pMpc. However, for higher mass haloes, star formation wouldn't be suppressed; the heating caused by the ionizing radiation would be considerably less (an order of magnitude) than that of the high-mass collapsing gas cloud \citep{ThoulWeinberg1996}. Therefore in the case of Lyman Break Galaxies (LBGs), which are generally accepted to have formed earlier and to be of larger mass than LAEs \citep{Kashikawa2007} the effects of star formation suppression due to ionizing photons from the quasar should be smaller, if not entirely negligible. Given this, if the hole around the center is indeed caused by negative feedback, then an LBG search in the same field should reveal an overdensity of LBGs in general and populate the inner 5~pMpc around the quasar in particular.

 Different quasar properties would influence the total observable suppression. This might explain the diversity of results within the literature, for example, a more luminous quasar would be able to ionize higher mass haloes. Another factor would be the age of the quasar and the light travel time. Quasars need time to ionize their surroundings. The longer the quasar has been turned on, the larger the suppression radius, up until $J_{21}=1$. Finally, the covering factor of the quasar ( along with the orientation of the host galaxy, or how off-centered the quasar is) would directly influence how many ionizing photons could escape the AGN and affect the IGM. Therefore we would expect a relationship between covering factor and overdensity. 

\subsubsection{Cosmic variance}
Another explanation for the lack of Ly-$\alpha$ emitters in the proximity of the quasar is cosmic variance; a chance arrangement of LAEs, without a physical cause. Indeed, this reason has been suggested by many studies to account for a lack of LAE concentrations around other high-redshift quasars \citep{Eduardo2013, Chiara2017,Ota2018}. Although in the previous section we concluded it is very unlikely that the underdensity around VIK 2348-3054 is caused by chance,  we cannot completely rule out this possibility. Interestingly \cite{Chen2020}, using an adaptive refinement tree code on a suite of simulations, showed that the lack of faint galaxies due to suppression of star formation from quasars was a less important factor on the cumulative luminosity function than simple field-to-field variation. They determined this by searching for LAEs in random 1 pMpc radii spheres in simulations. Their results suggest that cosmic variance is an effect which needs to be considered seriously in this context. Accounting for cosmic variance robustly would require further comparisons, on similar depths and FoVs, to other blank fields as well as wider field of view observations of other quasar fields. In contrast to star formation suppression, LBGs would be expected to show a similar pattern as LAEs around VIK J2348--3054 if the observed structure is a result of cosmic variance.

\subsubsection{Sensitivity}
As mentioned, suppression of Ly-$\alpha$ emission should occur in galaxies with halo masses less than $10^{9}M_{\odot}$. However, if a survey is unable to detect faint LAEs which fall into this regime, then it is possible that the faint LAEs do exist around the central quasar, but are just undetected. It is important to note that there have been confirmed detections of LAEs near quasars that LAEs near quasars have been spectroscopically confirmed using JWST \cite{Iani2024}. This suggests that follow up observations at much higher sensitivities, possibly using JWST, are needed. However, if sensitivity were the reason for the lack of detected LAEs around VIKJ2348--3054 then this would then raise the question as to why we don't detect any high-mass LAEs in the immediate vicinity of the quasar. So while sensitivity is an important component to consider, it does not fully explain our observed distribution of LAEs.

\subsubsection{Other possible physical mechanisms}
The lack of central LAEs near the quasar might be physically driven by processes other than star formation suppression by the quasar and cosmic variance.

A merging scenario was suggested by \cite{Eduardo2013}. In this model, the central quasar grows its mass through hierarchical mergers with surrounding haloes, creating an underdensity in its vicinity. This scenario would also potentially explain the quick SMBH growth experienced by high-redshift quasars and, at the same time, explains the apparent lack of the LAEs in the nearby vicinity as they would have merged with the host galaxy of the quasar.

On the other hand, \cite{Willott2005}, \cite{Kim2009}, and \cite{Chiara2017}, raised the possibility that candidates around the quasar might be dust obscured, thereby resulting in non-detections. However, explaining our results would require a gradient in the amount of dust as a function of the distance to the quasar, which seems unlikely. Submm interferometric surveys within the inner 5~pMpc might be able to confirm or deny this effect. Although, mapping such a large area with, e.g., ALMA, would be observationally expensive. For example, ALMA observations of VIK J2348--3054 by \cite{Venemans2016}, which cover an area of only 0.13 arcmin$^{2}$ area, found no companion galaxies. Recently, \cite{Li2023} searched for submm galaxies around $z \sim 6$ quasars---including VIK J2348--3054---using the Submillimeter Common-User Bolometre Array-2 (SCUBA-2) on the \textit{James Clerk Maxwell} Telescope, which has a much larger 15' field-of-view. This field of view translates to $\sim$ 5pMpc. \cite{Li2023} only found a single SMG around VIK J2348--3054, 0.5 pMpc away from the quasar, suggesting that dust obscuration may not be the cause of the central lack of LAEs.

Whilst these suggested mechanisms may be reasonable to explain the results of individual observations, they fail to account for the breadth of results throughout the literature---underdensities, field-densities, and overdensities. Considering this, we conclude that negative feedback is the most likely explanation of the observed results in this work.

\subsection{Comparison to other studies}
\begin{figure}
    \centering
    \includegraphics[width=0.5\textwidth]{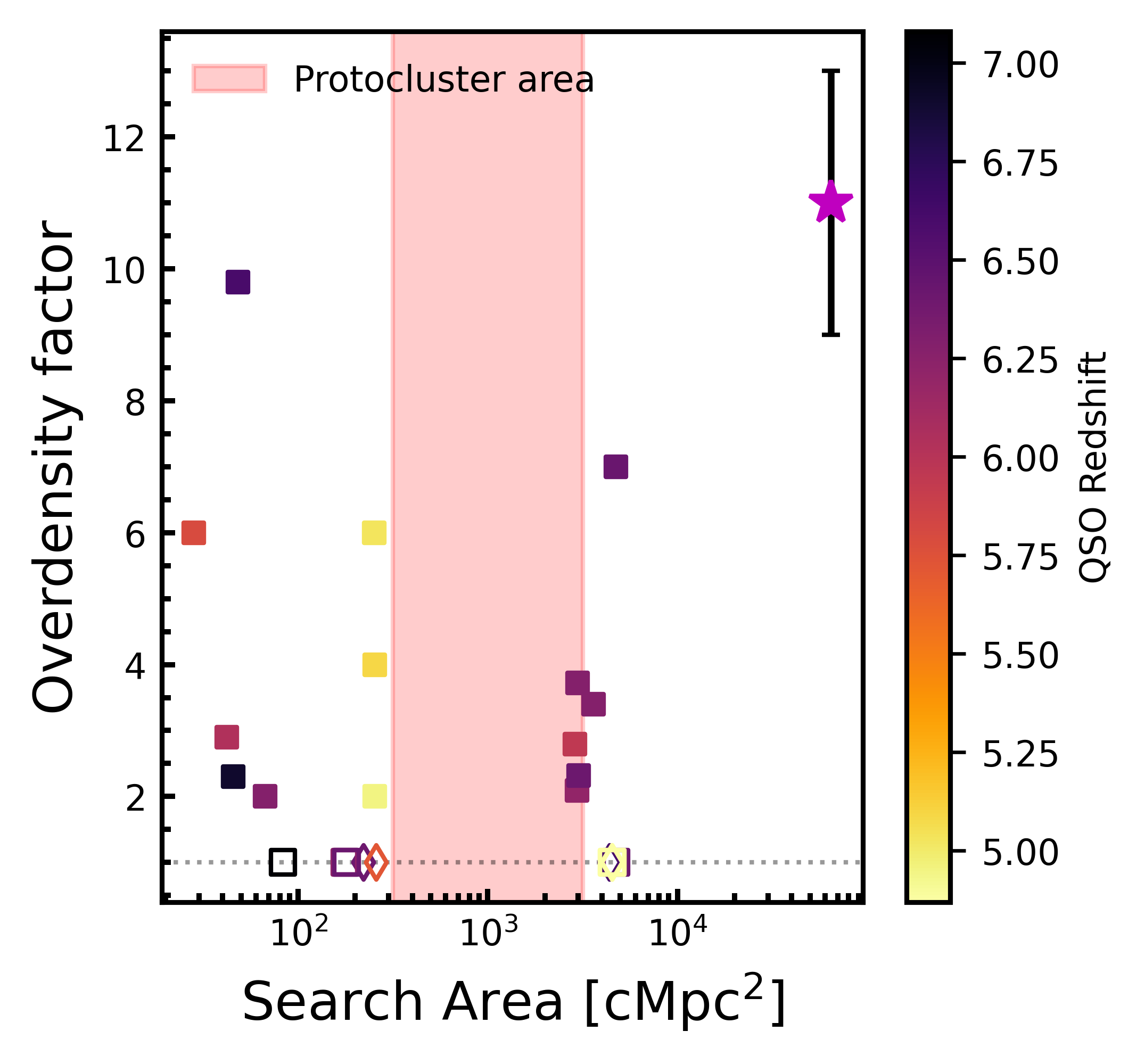}
    \caption{Summary plot of the various studies and results of looking for LAEs and LBGs around high-redshift quasars. Squares are Lyman-break searches and diamonds are Ly-$\alpha$ studies. Filled in values indicate that an overdensity was found, whilst not filled in represents that no overdensity was found. The purple star shows the overdensity of this study when compared to luminosity functions by \cite{Hu2019} and \cite{Wold2022}. The shaded pink region shows the area over which protoclusters are typically expected to extend according to \cite{Overzier2009}.}
    \label{fig:chiara_plot}
\end{figure}

The varying results throughout the literature have been difficult to constrain and many explanations have been given for the scatter. In Figure \ref{fig:chiara_plot} we summarize the results of previous studies that targeted 20 quasars in the redshift range $z=4.87$ to $z=7.08$, that searched for either LAEs or LBGs in their vicinity. Specifically, we show the search area and the value of $\delta$ determined by each study.

Of the 20 quasars only six were searched for neighboring LAEs. And of these six observations, none found overdensities. Furthermore there were no multiple LAE searches done on any of these quasars, meaning that only six LAE searches were done in total \citep{Eduardo2013, Chiara2017, Kikuta2017, goto2017, Ota2018}. On the other hand, 16 of the quasars had LBG surveys done, and some quasars were studied multiple times by different groups. In total there were 24 LBG searches done, of which 16 (67$\%$) found overdensities, whilst 8 (33$\%$) did not. The fact that many LAE searches don't detect overdensities, whilst at the same time, most LBG searches do, suggests that quasars may be suppressing star-formation in their environments. As we mentioned previously, the younger, lower mass LAEs would be more affected than the older, higher mass LBGs. Therefore, we would expect to find less overdensities when looking for LAEs. Alternatively, narrowband imaging might not have enough sensitivity if the equivalent widths are low.

 Several quasars have had both LAEs and LBGs searches done. VIKJ1030--0524 was investigated for LBGs by \cite{Morselli2014} and \cite{Balmaverde2017}. Later, \cite{Mignoli2020} performed spectroscopy using MUSE at VLT and spectroscopically identified the LBGs within their FoV, and found two LAE within 5 pMpc of the quasar. \cite{Mignoli2020} themselves note that the low number of detected LAEs close to the quasar may be hinting at negative feedback; confirming an overdensity in general but lacking in LAEs.
 
 \cite{Ota2018} searched for both LBGs and LAEs around VIK J0305--3150 and found, at low masses, a slight overdensity of LBGs and an underdensity of LAEs. Later, \cite{Champagne2023} reported a massive overdensity of LBGs very near the same quasar. Whilst narrowband searches for LAEs around the quasar have resulted in non-detections, spectroscopic observations, in particular with MUSE on the Very Large Telescope, identified a LAE very near to the quasar \citep{Farina2017}. Recently, \textit{James Webb} Space Telescope (JWST) slitless spectroscopy has identified 10 [OIII] emitting galaxies, which trace a filamentary structure around this quasar \citep{Wang2023}. These results highlight the importance that spectroscopic follow up can play in determining overdensities.
 
 \cite{Utsumi2010} and \cite{goto2017} both observed CFHQS J2329--0301 for LBGs and LAEs respectively, with Subaru/Suprime-Cam. Once again, the quasar had an overdensity of LBGs and an underdensity of LAEs.

\cite{Eduardo2013} and \cite{Chiara2017} searched for LAEs around ULAS J0203+0012 and PSO J215.1512–16.0417 respectively, and also performed Lyman-break selections around their respective targets. They found that the number of LBG candidates was consistent with blank fields. It is worth noting that both used the  FOcal Reducer/low
dispersion Spectrograph 2 (FORS2) on the Very Large Telescope (VLT) which has a FoV of 37 arcmin$^{2}$ and they did not use an optimum filter set up, which may have influenced results.

 Recently, \cite{Champagne2023} performed an LBG search around the target of our study for LAEs---VIK J2348--3054---using the Advanced Camera for Surveys (ACS) onboard the \textit{Hubble} Space Telescope (HST). Whilst they claim no LBG enhancement, they still find 10 LBG candidates within 0.3 pMpc, well within the quasar proximity zone (5 secure detections and 5 marginal detections). On the other hand, we do not find any LAEs within a much larger area around VIK J2348--3054. Once again, the lack of LAE and the detection of LBGs is expected, further hinting at star formation suppression. 

 There is also evidence at lower redshifts ($z\sim4$) which suggest a similar effect might be occurring. Specifically, \cite{GarciaVergara2017} used a custom set of narrowband filters which were optimized for identifying LBGs around six quasars; the QSO-LBG cross-correlation function showed an overdensity. Later, \cite{GarciaVergara2019} performed a similar work, but for LAEs around 17 quasars and again the QSO-LAE cross-correlation function found an overdensity. However, the LBG overdensity signal was stronger than the LAE one by a factor of 10 \citep[see the bottom panel of Figure 6 in][]{GarciaVergara2022}.
 
 The notion of LAE suppression due to the quasar seems to explain a lot of the results throughout the literature. However, at the same time, our study is the only one which has found an overdensity of LAEs around a high-redshift quasar ($z>5$) using the narrowband search technique. And is the largest area ever searched for either LBGs or LAEs (see the solid purple star in Figure \ref{fig:chiara_plot}). Moreover, had we observed VIK J2348--3054 with any other observational set up, on any of the other telescopes that were used in previous observations,  we would not have detected an overdensity. This would seem to suggest that the lack of overdensities of LAEs in previous studies might have been due to not probing a large enough area, and wider FoV studies would possibly have revealed overdense regions.

\section{Summary and Conclusion}\label{sec:conclusion}
In this paper we aimed to address important observational factors we believe have influenced previous studies which searched for overdensities around high-redshift quasars. Specifically, we searched for LAEs over an area of 3.26 deg$^2$, much larger than that probed by previous studies (see Figure \ref{fig:chiara_plot}), constituting the largest FoV searched for LAEs or LBGs around a high-redshift quasar to date. Additionally, we use a narrow-band that covers Ly-$\alpha$ at the systemic redshift of a quasar determined from [CII] observations, to avoid biases due to uncertain redshift estimates from, e.g., CIV and MgII \citep[see][]{Decarli2018}.

We found 38 LAE candidates around VIK J2348--3054 and applied our same selection criteria to the CDFS, images obtained with the same setup by \cite{Hu2019}, for comparison. After degrading the CDFS data to match our shallower depths, we found only two LAE candidates. Adjusting for the differences in the usable areas of both fields, this implies that VIK J2348--3054 resides in an environment that is overdense by a factor of $\delta = 10^{+5.3}_{-5.9}$. The space density of our candidates also sit well above the $z\sim 7$ luminosity functions of both \cite{Hu2019} and \cite{Wold2022} by a factor of $\delta$ = 4$\pm$0.5, further supporting the presence of a large overdensity.

Interestingly, we identified a distinct lack of LAEs in the immediate vicinity around the quasar itself, within 5.15 pMpc. Assuming that the surface density at distances further than  75 cMpc \citep{Overzier2009} as representative for the field, we find that the lack of LAEs withing 5.15 pMpc of the quasar has a probability of only 1.2$\%$ of occurring by chance, which is not enough to exclude cosmic variance as a cause. However, given that the underdensity is centered on the quasar itself, we take this to be a conservative estimate and while we cannot completely rule out cosmic variance, we conclude that this suppression is more likely due to a physical mechanism associated with the quasar, and propose it to be because of ionizing radiation which is inhibiting star formation in nearby ($< 5$~pMpc) galaxies. We calculate the isotropic UV intensity at the Lyman limit at 5~pMpc for VIK J2348--3054 and show that this would be sufficient to suppress star formation in low mass galaxies such as LAEs but wouldn't be enough to suppress it in larger galaxies like LBGs. It is therefore interesting that a recent small-FoV search for LBGs around VIK J2348--3054 found 10 LBG candidates within 0.3 pMpc \citep{Champagne2023}. 

As shown in Figure \ref{fig:chiara_plot}, there is a wide-spread of results for studies on the environmental density of high redshift quasars. Based on the results of our study, we potentially attribute this to a variety of factors. If the quasar is indeed suppressing star formation in its nearby vicinity then that suppression would be dependent on the luminosity, age, and covering factor of the quasar, and the combination of one or any of these factors could induce a spread of results. Alternatively, smaller FoV studies may be dominated by cosmic variance. They might also be probing areas of star-formation suppression, therefore adding to the confusion of whether quasars live in overdense environments.

Follow up observations of this field will be essential; LBG searches in particular, as well as spectroscopic follow up will help differentiate between cosmic variance and suppression of star formation.

This work has highlighted the importance of large-FoV studies and how not probing a large enough area might result in false non-detections of overdensities. It is therefore critical to perform large-FoV follow up surveys of other quasar fields, not only to cover regions much wider than those where star-formation might be suppressed by the quasar, but also to investigate the role that cosmic variance has on these kinds of surveys. It is also clear that future searches will require consistent methodologies to observationally confirm whether or not quasars live in overdense regions. Therefore, a large-FoV campaign to observe multiple high-redshift quasars would be invaluable.

\begin{acknowledgements}
We thank the anonymous referee for the valuable comments. Based on observations at Cerro Tololo Inter-American Observatory, NSF’s NOIRLab (NOIRLab Prop. ID 2021B-0905; PI: R. Assef), which is managed by the Association of Universities for Research in Astronomy (AURA) under a cooperative agreement with the National Science Foundation.

TSL, RJA, and AP acknowledge support from grant CONICYT + PCI + INSTITUTO MAX PLANCK DE ASTRONOMIA MPG190030. RJA acknowledge support from ANID BASAL project FB210003. RJA was supported by FONDECYT grant numbers 1191124 and 123171, and by the ANID BASAL project FB210003. CM acknowledges support from Fondecyt Iniciacion grant 11240336  and the ANID BASAL project FB210003. JXW acknowledges support from the science research grant from the China
Manned Space Project with No. CMS-CSST-2021-A07.
\end{acknowledgements}

\bibliographystyle{aa} % style aa.bst
\bibliography{trystan} % your references Yourfile.bib

\begin{thebibliography}{74}
\expandafter\ifx\csname natexlab\endcsname\relax\def\natexlab#1{#1}\fi

\bibitem[{{Abbott} {et~al.}(2021){Abbott}, {Adam{\'o}w}, {Aguena}, {Allam},
  {Amon}, {Annis}, {Avila}, {Bacon}, {Banerji}, {Bechtol}, {Becker},
  {Bernstein}, {Bertin}, {Bhargava}, {Bridle}, {Brooks}, {Burke}, {Carnero
  Rosell}, {Carrasco Kind}, {Carretero}, {Castander}, {Cawthon}, {Chang},
  {Choi}, {Conselice}, {Costanzi}, {Crocce}, {da Costa}, {Davis}, {De Vicente},
  {DeRose}, {Desai}, {Diehl}, {Dietrich}, {Drlica-Wagner}, {Eckert},
  {Elvin-Poole}, {Everett}, {Evrard}, {Ferrero}, {Fert{\'e}}, {Flaugher},
  {Fosalba}, {Friedel}, {Frieman}, {Garc{\'\i}a-Bellido}, {Gaztanaga},
  {Gelman}, {Gerdes}, {Giannantonio}, {Gill}, {Gruen}, {Gruendl}, {Gschwend},
  {Gutierrez}, {Hartley}, {Hinton}, {Hollowood}, {Honscheid}, {Huterer},
  {James}, {Jeltema}, {Johnson}, {Kent}, {Kron}, {Kuehn}, {Kuropatkin},
  {Lahav}, {Li}, {Lidman}, {Lin}, {MacCrann}, {Maia}, {Manning}, {Maloney},
  {March}, {Marshall}, {Martini}, {Melchior}, {Menanteau}, {Miquel}, {Morgan},
  {Myles}, {Neilsen}, {Ogando}, {Palmese}, {Paz-Chinch{\'o}n}, {Petravick},
  {Pieres}, {Plazas}, {Pond}, {Rodriguez-Monroy}, {Romer}, {Roodman}, {Rykoff},
  {Sako}, {Sanchez}, {Santiago}, {Scarpine}, {Serrano}, {Sevilla-Noarbe},
  {Smith}, {Smith}, {Soares-Santos}, {Suchyta}, {Swanson}, {Tarle}, {Thomas},
  {To}, {Tremblay}, {Troxel}, {Tucker}, {Turner}, {Varga}, {Walker},
  {Wechsler}, {Weller}, {Wester}, {Wilkinson}, {Yanny}, {Zhang}, {Nikutta},
  {Fitzpatrick}, {Jacques}, {Scott}, {Olsen}, {Huang}, {Herrera}, {Juneau},
  {Nidever}, {Weaver}, {Adean}, {Correia}, {de Freitas}, {Freitas},
  {Singulani}, {Vila-Verde}, \& {Linea Science Server}}]{DES2021}
{Abbott}, T.~M.~C., {Adam{\'o}w}, M., {Aguena}, M., {et~al.} 2021, \apjs, 255,
  20

\bibitem[{{Angulo} {et~al.}(2012){Angulo}, {Springel}, {White}, {Cole},
  {Jenkins}, {Baugh}, \& {Frenk}}]{Angulo2012}
{Angulo}, R.~E., {Springel}, V., {White}, S.~D.~M., {et~al.} 2012, \mnras, 425,
  2722

\bibitem[{{Ba{\~n}ados} {et~al.}(2013){Ba{\~n}ados}, {Venemans}, {Walter},
  {Kurk}, {Overzier}, \& {Ouchi}}]{Eduardo2013}
{Ba{\~n}ados}, E., {Venemans}, B., {Walter}, F., {et~al.} 2013, \apj, 773, 178

\bibitem[{{Balmaverde} {et~al.}(2017){Balmaverde}, {Gilli}, {Mignoli},
  {Bolzonella}, {Brusa}, {Cappelluti}, {Comastri}, {Sani}, {Vanzella},
  {Vignali}, {Vito}, \& {Zamorani}}]{Balmaverde2017}
{Balmaverde}, B., {Gilli}, R., {Mignoli}, M., {et~al.} 2017, \aap, 606, A23

\bibitem[{{Bertin} \& {Arnouts}(1996)}]{sextractor}
{Bertin}, E. \& {Arnouts}, S. 1996, \aaps, 117, 393

\bibitem[{{Bosman} {et~al.}(2022){Bosman}, {Davies}, {Becker}, {Keating},
  {Davies}, {Zhu}, {Eilers}, {D'Odorico}, {Bian}, {Bischetti}, {Cristiani},
  {Fan}, {Farina}, {Haehnelt}, {Hennawi}, {Kulkarni}, {Mesinger}, {Meyer},
  {Onoue}, {Pallottini}, {Qin}, {Ryan-Weber}, {Schindler}, {Walter}, {Wang}, \&
  {Yang}}]{Bosman2022}
{Bosman}, S. E.~I., {Davies}, F.~B., {Becker}, G.~D., {et~al.} 2022, \mnras,
  514, 55

\bibitem[{{Capak} {et~al.}(2011){Capak}, {Riechers}, {Scoville}, {Carilli},
  {Cox}, {Neri}, {Robertson}, {Salvato}, {Schinnerer}, {Yan}, {Wilson}, {Yun},
  {Civano}, {Elvis}, {Karim}, {Mobasher}, \& {Staguhn}}]{Capak2011}
{Capak}, P.~L., {Riechers}, D., {Scoville}, N.~Z., {et~al.} 2011, \nat, 470,
  233

\bibitem[{{Chambers} {et~al.}(2016){Chambers}, {Magnier}, {Metcalfe},
  {Flewelling}, {Huber}, {Waters}, {Denneau}, {Draper}, {Farrow}, {Finkbeiner},
  {Holmberg}, {Koppenhoefer}, {Price}, {Rest}, {Saglia}, {Schlafly}, {Smartt},
  {Sweeney}, {Wainscoat}, {Burgett}, {Chastel}, {Grav}, {Heasley}, {Hodapp},
  {Jedicke}, {Kaiser}, {Kudritzki}, {Luppino}, {Lupton}, {Monet}, {Morgan},
  {Onaka}, {Shiao}, {Stubbs}, {Tonry}, {White}, {Ba{\~n}ados}, {Bell},
  {Bender}, {Bernard}, {Boegner}, {Boffi}, {Botticella}, {Calamida},
  {Casertano}, {Chen}, {Chen}, {Cole}, {Deacon}, {Frenk}, {Fitzsimmons},
  {Gezari}, {Gibbs}, {Goessl}, {Goggia}, {Gourgue}, {Goldman}, {Grant},
  {Grebel}, {Hambly}, {Hasinger}, {Heavens}, {Heckman}, {Henderson}, {Henning},
  {Holman}, {Hopp}, {Ip}, {Isani}, {Jackson}, {Keyes}, {Koekemoer}, {Kotak},
  {Le}, {Liska}, {Long}, {Lucey}, {Liu}, {Martin}, {Masci}, {McLean}, {Mindel},
  {Misra}, {Morganson}, {Murphy}, {Obaika}, {Narayan}, {Nieto-Santisteban},
  {Norberg}, {Peacock}, {Pier}, {Postman}, {Primak}, {Rae}, {Rai}, {Riess},
  {Riffeser}, {Rix}, {R{\"o}ser}, {Russel}, {Rutz}, {Schilbach}, {Schultz},
  {Scolnic}, {Strolger}, {Szalay}, {Seitz}, {Small}, {Smith}, {Soderblom},
  {Taylor}, {Thomson}, {Taylor}, {Thakar}, {Thiel}, {Thilker}, {Unger},
  {Urata}, {Valenti}, {Wagner}, {Walder}, {Walter}, {Watters}, {Werner},
  {Wood-Vasey}, \& {Wyse}}]{PANSTARS1}
{Chambers}, K.~C., {Magnier}, E.~A., {Metcalfe}, N., {et~al.} 2016, arXiv
  e-prints, arXiv:1612.05560

\bibitem[{{Champagne} {et~al.}(2023){Champagne}, {Casey}, {Finkelstein},
  {Bagley}, {Cooper}, {Larson}, {Long}, \& {Wang}}]{Champagne2023}
{Champagne}, J.~B., {Casey}, C.~M., {Finkelstein}, S.~L., {et~al.} 2023, \apj,
  952, 99

\bibitem[{{Chen}(2020)}]{Chen2020}
{Chen}, H. 2020, \apj, 893, 165

\bibitem[{{Chiang} {et~al.}(2017){Chiang}, {Overzier}, {Gebhardt}, \&
  {Henriques}}]{Chiang2017}
{Chiang}, Y.-K., {Overzier}, R.~A., {Gebhardt}, K., \& {Henriques}, B. 2017,
  \apjl, 844, L23

\bibitem[{{Costa} {et~al.}(2014){Costa}, {Sijacki}, {Trenti}, \&
  {Haehnelt}}]{Costa2014}
{Costa}, T., {Sijacki}, D., {Trenti}, M., \& {Haehnelt}, M.~G. 2014, \mnras,
  439, 2146

\bibitem[{{Decarli} {et~al.}(2019){Decarli}, {Mignoli}, {Gilli}, {Balmaverde},
  {Brusa}, {Cappelluti}, {Comastri}, {Nanni}, {Peca}, {Pensabene}, {Vanzella},
  \& {Vignali}}]{Decarli2019}
{Decarli}, R., {Mignoli}, M., {Gilli}, R., {et~al.} 2019, \aap, 631, L10

\bibitem[{{Decarli} {et~al.}(2018){Decarli}, {Walter}, {Venemans},
  {Ba{\~n}ados}, {Bertoldi}, {Carilli}, {Fan}, {Farina}, {Mazzucchelli},
  {Riechers}, {Rix}, {Strauss}, {Wang}, \& {Yang}}]{Decarli2018}
{Decarli}, R., {Walter}, F., {Venemans}, B.~P., {et~al.} 2018, \apj, 854, 97

\bibitem[{{D{\'\i}az-Santos} {et~al.}(2021){D{\'\i}az-Santos}, {Assef},
  {Eisenhardt}, {Jun}, {Jones}, {Blain}, {Stern}, {Aravena}, {Tsai}, {Lake},
  {Wu}, \& {Gonz{\'a}lez-L{\'o}pez}}]{Tanio2021}
{D{\'\i}az-Santos}, T., {Assef}, R.~J., {Eisenhardt}, P. R.~M., {et~al.} 2021,
  \aap, 654, A37

\bibitem[{{Eilers} {et~al.}(2023){Eilers}, {Simcoe}, {Yue}, {Mackenzie},
  {Matthee}, {{\v{D}}urov{\v{c}}{\'\i}kov{\'a}}, {Kashino}, {Bordoloi}, \&
  {Lilly}}]{Eilers2023}
{Eilers}, A.-C., {Simcoe}, R.~A., {Yue}, M., {et~al.} 2023, \apj, 950, 68

\bibitem[{{Fan} {et~al.}(2023){Fan}, {Ba{\~n}ados}, \&
  {Simcoe}}]{EduardoReviewQSO}
{Fan}, X., {Ba{\~n}ados}, E., \& {Simcoe}, R.~A. 2023, \araa, 61, 373

\bibitem[{{Fan} {et~al.}(2006){Fan}, {Strauss}, {Becker}, {White}, {Gunn},
  {Knapp}, {Richards}, {Schneider}, {Brinkmann}, \& {Fukugita}}]{Fan2006}
{Fan}, X., {Strauss}, M.~A., {Becker}, R.~H., {et~al.} 2006, \aj, 132, 117

\bibitem[{{Fan} {et~al.}(2001){Fan}, {Strauss}, {Richards}, {Newman}, {Becker},
  {Schneider}, {Gunn}, {Davis}, {White}, {Lupton}, {Anderson}, {Annis},
  {Bahcall}, {Brunner}, {Csabai}, {Doi}, {Fukugita}, {Hennessy}, {Hindsley},
  {Ivezi{\'c}}, {Knapp}, {McKay}, {Munn}, {Pier}, {Szalay}, \&
  {York}}]{Fan2001}
{Fan}, X., {Strauss}, M.~A., {Richards}, G.~T., {et~al.} 2001, \aj, 121, 31

\bibitem[{{Farina} {et~al.}(2022){Farina}, {Schindler}, {Walter},
  {Ba{\~n}ados}, {Davies}, {Decarli}, {Eilers}, {Fan}, {Hennawi},
  {Mazzucchelli}, {Meyer}, {Trakhtenbrot}, {Volonteri}, {Wang}, {Worseck},
  {Yang}, {Gutcke}, {Venemans}, {Bosman}, {Costa}, {De Rosa}, {Drake}, \&
  {Onoue}}]{Farina2022}
{Farina}, E.~P., {Schindler}, J.-T., {Walter}, F., {et~al.} 2022, \apj, 941,
  106

\bibitem[{{Farina} {et~al.}(2017){Farina}, {Venemans}, {Decarli}, {Hennawi},
  {Walter}, {Ba{\~n}ados}, {Mazzucchelli}, {Cantalupo}, {Arrigoni-Battaia}, \&
  {McGreer}}]{Farina2017}
{Farina}, E.~P., {Venemans}, B.~P., {Decarli}, R., {et~al.} 2017, \apj, 848, 78

\bibitem[{{Flaugher} {et~al.}(2015){Flaugher}, {Diehl}, {Honscheid}, {Abbott},
  {Alvarez}, {Angstadt}, {Annis}, {Antonik}, {Ballester}, {Beaufore},
  {Bernstein}, {Bernstein}, {Bigelow}, {Bonati}, {Boprie}, {Brooks},
  {Buckley-Geer}, {Campa}, {Cardiel-Sas}, {Castander}, {Castilla}, {Cease},
  {Cela-Ruiz}, {Chappa}, {Chi}, {Cooper}, {da Costa}, {Dede}, {Derylo},
  {DePoy}, {de Vicente}, {Doel}, {Drlica-Wagner}, {Eiting}, {Elliott}, {Emes},
  {Estrada}, {Fausti Neto}, {Finley}, {Flores}, {Frieman}, {Gerdes},
  {Gladders}, {Gregory}, {Gutierrez}, {Hao}, {Holland}, {Holm}, {Huffman},
  {Jackson}, {James}, {Jonas}, {Karcher}, {Karliner}, {Kent}, {Kessler},
  {Kozlovsky}, {Kron}, {Kubik}, {Kuehn}, {Kuhlmann}, {Kuk}, {Lahav}, {Lathrop},
  {Lee}, {Levi}, {Lewis}, {Li}, {Mandrichenko}, {Marshall}, {Martinez},
  {Merritt}, {Miquel}, {Mu{\~n}oz}, {Neilsen}, {Nichol}, {Nord}, {Ogando},
  {Olsen}, {Palaio}, {Patton}, {Peoples}, {Plazas}, {Rauch}, {Reil}, {Rheault},
  {Roe}, {Rogers}, {Roodman}, {Sanchez}, {Scarpine}, {Schindler}, {Schmidt},
  {Schmitt}, {Schubnell}, {Schultz}, {Schurter}, {Scott}, {Serrano}, {Shaw},
  {Smith}, {Soares-Santos}, {Stefanik}, {Stuermer}, {Suchyta}, {Sypniewski},
  {Tarle}, {Thaler}, {Tighe}, {Tran}, {Tucker}, {Walker}, {Wang}, {Watson},
  {Weaverdyck}, {Wester}, {Woods}, {Yanny}, \& {DES
  Collaboration}}]{Flaugher2015}
{Flaugher}, B., {Diehl}, H.~T., {Honscheid}, K., {et~al.} 2015, \aj, 150, 150

\bibitem[{{Gaia Collaboration} {et~al.}(2018){Gaia Collaboration}, {Brown},
  {Vallenari}, {Prusti}, {de Bruijne}, {Babusiaux}, {Bailer-Jones}, {Biermann},
  {Evans}, {Eyer}, {Jansen}, {Jordi}, {Klioner}, {Lammers}, {Lindegren},
  {Luri}, {Mignard}, {Panem}, {Pourbaix}, {Randich}, {Sartoretti}, {Siddiqui},
  {Soubiran}, {van Leeuwen}, {Walton}, {Arenou}, {Bastian}, {Cropper},
  {Drimmel}, {Katz}, {Lattanzi}, {Bakker}, {Cacciari}, {Casta{\~n}eda},
  {Chaoul}, {Cheek}, {De Angeli}, {Fabricius}, {Guerra}, {Holl}, {Masana},
  {Messineo}, {Mowlavi}, {Nienartowicz}, {Panuzzo}, {Portell}, {Riello},
  {Seabroke}, {Tanga}, {Th{\'e}venin}, {Gracia-Abril}, {Comoretto},
  {Garcia-Reinaldos}, {Teyssier}, {Altmann}, {Andrae}, {Audard},
  {Bellas-Velidis}, {Benson}, {Berthier}, {Blomme}, {Burgess}, {Busso},
  {Carry}, {Cellino}, {Clementini}, {Clotet}, {Creevey}, {Davidson}, {De
  Ridder}, {Delchambre}, {Dell'Oro}, {Ducourant},
  {Fern{\'a}ndez-Hern{\'a}ndez}, {Fouesneau}, {Fr{\'e}mat}, {Galluccio},
  {Garc{\'\i}a-Torres}, {Gonz{\'a}lez-N{\'u}{\~n}ez}, {Gonz{\'a}lez-Vidal},
  {Gosset}, {Guy}, {Halbwachs}, {Hambly}, {Harrison}, {Hern{\'a}ndez},
  {Hestroffer}, {Hodgkin}, {Hutton}, {Jasniewicz}, {Jean-Antoine-Piccolo},
  {Jordan}, {Korn}, {Krone-Martins}, {Lanzafame}, {Lebzelter}, {L{\"o}ffler},
  {Manteiga}, {Marrese}, {Mart{\'\i}n-Fleitas}, {Moitinho}, {Mora}, {Muinonen},
  {Osinde}, {Pancino}, {Pauwels}, {Petit}, {Recio-Blanco}, {Richards},
  {Rimoldini}, {Robin}, {Sarro}, {Siopis}, {Smith}, {Sozzetti}, {S{\"u}veges},
  {Torra}, {van Reeven}, {Abbas}, {Abreu Aramburu}, {Accart}, {Aerts},
  {Altavilla}, {{\'A}lvarez}, {Alvarez}, {Alves}, {Anderson}, {Andrei},
  {Anglada Varela}, {Antiche}, {Antoja}, {Arcay}, {Astraatmadja}, {Bach},
  {Baker}, {Balaguer-N{\'u}{\~n}ez}, {Balm}, {Barache}, {Barata}, {Barbato},
  {Barblan}, {Barklem}, {Barrado}, {Barros}, {Barstow}, {Bartholom{\'e}
  Mu{\~n}oz}, {Bassilana}, {Becciani}, {Bellazzini}, {Berihuete}, {Bertone},
  {Bianchi}, {Bienaym{\'e}}, {Blanco-Cuaresma}, {Boch}, {Boeche}, {Bombrun},
  {Borrachero}, {Bossini}, {Bouquillon}, {Bourda}, {Bragaglia}, {Bramante},
  {Breddels}, {Bressan}, {Brouillet}, {Br{\"u}semeister}, {Brugaletta},
  {Bucciarelli}, {Burlacu}, {Busonero}, {Butkevich}, {Buzzi}, {Caffau},
  {Cancelliere}, {Cannizzaro}, {Cantat-Gaudin}, {Carballo}, {Carlucci},
  {Carrasco}, {Casamiquela}, {Castellani}, {Castro-Ginard}, {Charlot},
  {Chemin}, {Chiavassa}, {Cocozza}, {Costigan}, {Cowell}, {Crifo}, {Crosta},
  {Crowley}, {Cuypers}, {Dafonte}, {Damerdji}, {Dapergolas}, {David}, {David},
  {de Laverny}, {De Luise}, {De March}, {de Martino}, {de Souza}, {de Torres},
  {Debosscher}, {del Pozo}, {Delbo}, {Delgado}, {Delgado}, {Di Matteo},
  {Diakite}, {Diener}, {Distefano}, {Dolding}, {Drazinos}, {Dur{\'a}n},
  {Edvardsson}, {Enke}, {Eriksson}, {Esquej}, {Eynard Bontemps}, {Fabre},
  {Fabrizio}, {Faigler}, {Falc{\~a}o}, {Farr{\`a}s Casas}, {Federici},
  {Fedorets}, {Fernique}, {Figueras}, {Filippi}, {Findeisen}, {Fonti},
  {Fraile}, {Fraser}, {Fr{\'e}zouls}, {Gai}, {Galleti}, {Garabato},
  {Garc{\'\i}a-Sedano}, {Garofalo}, {Garralda}, {Gavel}, {Gavras}, {Gerssen},
  {Geyer}, {Giacobbe}, {Gilmore}, {Girona}, {Giuffrida}, {Glass}, {Gomes},
  {Granvik}, {Gueguen}, {Guerrier}, {Guiraud}, {Guti{\'e}rrez-S{\'a}nchez},
  {Haigron}, {Hatzidimitriou}, {Hauser}, {Haywood}, {Heiter}, {Helmi}, {Heu},
  {Hilger}, {Hobbs}, {Hofmann}, {Holland}, {Huckle}, {Hypki}, {Icardi},
  {Jan{\ss}en}, {Jevardat de Fombelle}, {Jonker}, {Juh{\'a}sz}, {Julbe},
  {Karampelas}, {Kewley}, {Klar}, {Kochoska}, {Kohley}, {Kolenberg},
  {Kontizas}, {Kontizas}, {Koposov}, {Kordopatis}, {Kostrzewa-Rutkowska},
  {Koubsky}, {Lambert}, {Lanza}, {Lasne}, {Lavigne}, {Le Fustec}, {Le
  Poncin-Lafitte}, {Lebreton}, {Leccia}, {Leclerc}, {Lecoeur-Taibi},
  {Lenhardt}, {Leroux}, {Liao}, {Licata}, {Lindstr{\o}m}, {Lister}, {Livanou},
  {Lobel}, {L{\'o}pez}, {Managau}, {Mann}, {Mantelet}, {Marchal}, {Marchant},
  {Marconi}, {Marinoni}, {Marschalk{\'o}}, {Marshall}, {Martino}, {Marton},
  {Mary}, {Massari}, {Matijevi{\v{c}}}, {Mazeh}, {McMillan}, {Messina},
  {Michalik}, {Millar}, {Molina}, {Molinaro}, {Moln{\'a}r}, {Montegriffo},
  {Mor}, {Morbidelli}, {Morel}, {Morris}, {Mulone}, {Muraveva}, {Musella},
  {Nelemans}, {Nicastro}, {Noval}, {O'Mullane}, {Ord{\'e}novic},
  {Ord{\'o}{\~n}ez-Blanco}, {Osborne}, {Pagani}, {Pagano}, {Pailler},
  {Palacin}, {Palaversa}, {Panahi}, {Pawlak}, {Piersimoni}, {Pineau}, {Plachy},
  {Plum}, {Poggio}, {Poujoulet}, {Pr{\v{s}}a}, {Pulone}, {Racero}, {Ragaini},
  {Rambaux}, {Ramos-Lerate}, {Regibo}, {Reyl{\'e}}, {Riclet}, {Ripepi}, {Riva},
  {Rivard}, {Rixon}, {Roegiers}, {Roelens}, {Romero-G{\'o}mez}, {Rowell},
  {Royer}, {Ruiz-Dern}, {Sadowski}, {Sagrist{\`a} Sell{\'e}s}, {Sahlmann},
  {Salgado}, {Salguero}, {Sanna}, {Santana-Ros}, {Sarasso}, {Savietto},
  {Schultheis}, {Sciacca}, {Segol}, {Segovia}, {S{\'e}gransan}, {Shih},
  {Siltala}, {Silva}, {Smart}, {Smith}, {Solano}, {Solitro}, {Sordo}, {Soria
  Nieto}, {Souchay}, {Spagna}, {Spoto}, {Stampa}, {Steele},
  {Steidelm{\"u}ller}, {Stephenson}, {Stoev}, {Suess}, {Surdej}, {Szabados},
  {Szegedi-Elek}, {Tapiador}, {Taris}, {Tauran}, {Taylor}, {Teixeira},
  {Terrett}, {Teyssandier}, {Thuillot}, {Titarenko}, {Torra Clotet}, {Turon},
  {Ulla}, {Utrilla}, {Uzzi}, {Vaillant}, {Valentini}, {Valette}, {van Elteren},
  {Van Hemelryck}, {van Leeuwen}, {Vaschetto}, {Vecchiato}, {Veljanoski},
  {Viala}, {Vicente}, {Vogt}, {von Essen}, {Voss}, {Votruba}, {Voutsinas},
  {Walmsley}, {Weiler}, {Wertz}, {Wevers}, {Wyrzykowski}, {Yoldas},
  {{\v{Z}}erjal}, {Ziaeepour}, {Zorec}, {Zschocke}, {Zucker}, {Zurbach}, \&
  {Zwitter}}]{GAIA2018}
{Gaia Collaboration}, {Brown}, A.~G.~A., {Vallenari}, A., {et~al.} 2018, \aap,
  616, A1

\bibitem[{{Garc{\'\i}a-Vergara} {et~al.}(2019){Garc{\'\i}a-Vergara}, {Hennawi},
  {Barrientos}, \& {Arrigoni Battaia}}]{GarciaVergara2019}
{Garc{\'\i}a-Vergara}, C., {Hennawi}, J.~F., {Barrientos}, L.~F., \& {Arrigoni
  Battaia}, F. 2019, \apj, 886, 79

\bibitem[{{Garc{\'\i}a-Vergara} {et~al.}(2017){Garc{\'\i}a-Vergara}, {Hennawi},
  {Barrientos}, \& {Rix}}]{GarciaVergara2017}
{Garc{\'\i}a-Vergara}, C., {Hennawi}, J.~F., {Barrientos}, L.~F., \& {Rix},
  H.-W. 2017, \apj, 848, 7

\bibitem[{{Garc{\'\i}a-Vergara} {et~al.}(2022){Garc{\'\i}a-Vergara}, {Rybak},
  {Hodge}, {Hennawi}, {Decarli}, {Gonz{\'a}lez-L{\'o}pez}, {Arrigoni-Battaia},
  {Aravena}, \& {Farina}}]{GarciaVergara2022}
{Garc{\'\i}a-Vergara}, C., {Rybak}, M., {Hodge}, J., {et~al.} 2022, \apj, 927,
  65

\bibitem[{{Gehrels}(1986)}]{SmallNumberStatistics}
{Gehrels}, N. 1986, \apj, 303, 336

\bibitem[{{Goto} {et~al.}(2017){Goto}, {Utsumi}, {Kikuta}, {Miyazaki}, {Shiki},
  \& {Hashimoto}}]{goto2017}
{Goto}, T., {Utsumi}, Y., {Kikuta}, S., {et~al.} 2017, \mnras, 470, L117

\bibitem[{{Hu} {et~al.}(2019){Hu}, {Wang}, {Zheng}, {Malhotra}, {Rhoads},
  {Infante}, {Barrientos}, {Yang}, {Jiang}, {Kang}, {Perez}, {Wold}, {Hibon},
  {Jiang}, {Khostovan}, {Valdes}, {Walker}, {Galaz}, {Coughlin}, {Harish},
  {Kong}, {Pharo}, \& {Zheng}}]{Hu2019}
{Hu}, W., {Wang}, J., {Zheng}, Z.-Y., {et~al.} 2019, \apj, 886, 90

\bibitem[{{Husband} {et~al.}(2013){Husband}, {Bremer}, {Stanway}, {Davies},
  {Lehnert}, \& {Douglas}}]{Husband2013}
{Husband}, K., {Bremer}, M.~N., {Stanway}, E.~R., {et~al.} 2013, \mnras, 432,
  2869

\bibitem[{{Iani} {et~al.}(2024){Iani}, {Caputi}, {Rinaldi}, {Annunziatella},
  {Boogaard}, {{\"O}stlin}, {Costantin}, {Gillman}, {P{\'e}rez-Gonz{\'a}lez},
  {Colina}, {Greve}, {Wright}, {Alonso-Herrero}, {{\'A}lvarez-M{\'a}rquez},
  {Bik}, {Bosman}, {Crespo G{\'o}mez}, {Eckart}, {Hjorth}, {Jermann},
  {Labiano}, {Langeroodi}, {Melinder}, {Moutard}, {Pei{\ss}ker}, {Pye},
  {Tikkanen}, {van der Werf}, {Walter}, {Henning}, {Lagage}, \& {van
  Dishoeck}}]{Iani2024}
{Iani}, E., {Caputi}, K.~I., {Rinaldi}, P., {et~al.} 2024, \apj, 963, 97

\bibitem[{{Inayoshi} {et~al.}(2020){Inayoshi}, {Visbal}, \&
  {Haiman}}]{Inayoshi2020}
{Inayoshi}, K., {Visbal}, E., \& {Haiman}, Z. 2020, \araa, 58, 27

\bibitem[{{Kashikawa} {et~al.}(2007){Kashikawa}, {Kitayama}, {Doi}, {Misawa},
  {Komiyama}, \& {Ota}}]{Kashikawa2007}
{Kashikawa}, N., {Kitayama}, T., {Doi}, M., {et~al.} 2007, \apj, 663, 765

\bibitem[{{Kennicutt} \& {Evans}(2012)}]{Kennicutt2012}
{Kennicutt}, R.~C. \& {Evans}, N.~J. 2012, \araa, 50, 531

\bibitem[{{Khostovan} {et~al.}(2020){Khostovan}, {Malhotra}, {Rhoads}, {Jiang},
  {Wang}, {Wold}, {Zheng}, {Barrientos}, {Coughlin}, {Harish}, {Hu}, {Infante},
  {Perez}, {Pharo}, {Valdes}, {Walker}, \& {Yang}}]{Khostovan2020}
{Khostovan}, A.~A., {Malhotra}, S., {Rhoads}, J.~E., {et~al.} 2020, \mnras,
  493, 3966

\bibitem[{{Kikuta} {et~al.}(2017){Kikuta}, {Imanishi}, {Matsuoka}, {Matsuda},
  {Shimasaku}, \& {Nakata}}]{Kikuta2017}
{Kikuta}, S., {Imanishi}, M., {Matsuoka}, Y., {et~al.} 2017, \apj, 841, 128

\bibitem[{{Kim} {et~al.}(2009){Kim}, {Stiavelli}, {Trenti}, {Pavlovsky},
  {Djorgovski}, {Scarlata}, {Stern}, {Mahabal}, {Thompson}, {Dickinson},
  {Panagia}, \& {Meylan}}]{Kim2009}
{Kim}, S., {Stiavelli}, M., {Trenti}, M., {et~al.} 2009, \apj, 695, 809

\bibitem[{{Li} {et~al.}(2023){Li}, {Wang}, {Fan}, {Wu}, {Jiang}, {Ba{\~n}ados},
  {Venemans}, {Shao}, {Li}, {Wagg}, {Decarli}, {Mazzucchelli}, {Omont},
  {Bertoldi}, {Johnson}, {Conselice}, \& {Zhang}}]{Li2023}
{Li}, Q., {Wang}, R., {Fan}, X., {et~al.} 2023, \apj, 954, 174

\bibitem[{{Madau}(1995)}]{Madau1995}
{Madau}, P. 1995, \apj, 441, 18

\bibitem[{{Malhotra} \& {Rhoads}(2002)}]{Malhotra2002}
{Malhotra}, S. \& {Rhoads}, J.~E. 2002, \apjl, 565, L71

\bibitem[{{Matthee} {et~al.}(2015){Matthee}, {Sobral}, {Santos},
  {R{\"o}ttgering}, {Darvish}, \& {Mobasher}}]{matthee2015}
{Matthee}, J., {Sobral}, D., {Santos}, S., {et~al.} 2015, \mnras, 451, 400

\bibitem[{{Mazzucchelli} {et~al.}(2017{\natexlab{a}}){Mazzucchelli},
  {Ba{\~n}ados}, {Decarli}, {Farina}, {Venemans}, {Walter}, \&
  {Overzier}}]{Chiara2017}
{Mazzucchelli}, C., {Ba{\~n}ados}, E., {Decarli}, R., {et~al.}
  2017{\natexlab{a}}, \apj, 834, 83

\bibitem[{{Mazzucchelli} {et~al.}(2017{\natexlab{b}}){Mazzucchelli},
  {Ba{\~n}ados}, {Venemans}, {Decarli}, {Farina}, {Walter}, {Eilers}, {Rix},
  {Simcoe}, {Stern}, {Fan}, {Schlafly}, {De Rosa}, {Hennawi}, {Chambers},
  {Greiner}, {Burgett}, {Draper}, {Kaiser}, {Kudritzki}, {Magnier}, {Metcalfe},
  {Waters}, \& {Wainscoat}}]{Chiara2017b}
{Mazzucchelli}, C., {Ba{\~n}ados}, E., {Venemans}, B.~P., {et~al.}
  2017{\natexlab{b}}, \apj, 849, 91

\bibitem[{{Mazzucchelli} {et~al.}(2023){Mazzucchelli}, {Bischetti},
  {D'Odorico}, {Feruglio}, {Schindler}, {Onoue}, {Ba{\~n}ados}, {Becker},
  {Bian}, {Carniani}, {Decarli}, {Eilers}, {Farina}, {Gallerani}, {Lai},
  {Meyer}, {Rojas-Ruiz}, {Satyavolu}, {Venemans}, {Wang}, {Yang}, \&
  {Zhu}}]{Chiara2023}
{Mazzucchelli}, C., {Bischetti}, M., {D'Odorico}, V., {et~al.} 2023, \aap, 676,
  A71

\bibitem[{{Mignoli} {et~al.}(2020){Mignoli}, {Gilli}, {Decarli}, {Vanzella},
  {Balmaverde}, {Cappelluti}, {Cassar{\`a}}, {Comastri}, {Cusano}, {Iwasawa},
  {Marchesi}, {Prandoni}, {Vignali}, {Vito}, {Zamorani}, {Chiaberge}, \&
  {Norman}}]{Mignoli2020}
{Mignoli}, M., {Gilli}, R., {Decarli}, R., {et~al.} 2020, \aap, 642, L1

\bibitem[{{Morselli} {et~al.}(2014){Morselli}, {Mignoli}, {Gilli}, {Vignali},
  {Comastri}, {Sani}, {Cappelluti}, {Zamorani}, {Brusa}, {Gallozzi}, \&
  {Vanzella}}]{Morselli2014}
{Morselli}, L., {Mignoli}, M., {Gilli}, R., {et~al.} 2014, \aap, 568, A1

\bibitem[{{Osterbrock}(1989)}]{Osterbrock1989}
{Osterbrock}, D.~E. 1989, {Astrophysics of gaseous nebulae and active galactic
  nuclei}

\bibitem[{{Ota} {et~al.}(2018){Ota}, {Venemans}, {Taniguchi}, {Kashikawa},
  {Nakata}, {Harikane}, {Ba{\~n}ados}, {Overzier}, {Riechers}, {Walter},
  {Toshikawa}, {Shibuya}, \& {Jiang}}]{Ota2018}
{Ota}, K., {Venemans}, B.~P., {Taniguchi}, Y., {et~al.} 2018, \apj, 856, 109

\bibitem[{{Ouchi} {et~al.}(2020){Ouchi}, {Ono}, \& {Shibuya}}]{Ouchi2020}
{Ouchi}, M., {Ono}, Y., \& {Shibuya}, T. 2020, \araa, 58, 617

\bibitem[{{Overzier} {et~al.}(2009){Overzier}, {Guo}, {Kauffmann}, {De Lucia},
  {Bouwens}, \& {Lemson}}]{Overzier2009}
{Overzier}, R.~A., {Guo}, Q., {Kauffmann}, G., {et~al.} 2009, \mnras, 394, 577

\bibitem[{{Polletta} {et~al.}(2007){Polletta}, {Tajer}, {Maraschi},
  {Trinchieri}, {Lonsdale}, {Chiappetti}, {Andreon}, {Pierre}, {Le F{\`e}vre},
  {Zamorani}, {Maccagni}, {Garcet}, {Surdej}, {Franceschini}, {Alloin},
  {Shupe}, {Surace}, {Fang}, {Rowan-Robinson}, {Smith}, \&
  {Tresse}}]{Polletta2007}
{Polletta}, M., {Tajer}, M., {Maraschi}, L., {et~al.} 2007, \apj, 663, 81

\bibitem[{{Schindler} {et~al.}(2020){Schindler}, {Farina}, {Ba{\~n}ados},
  {Eilers}, {Hennawi}, {Onoue}, {Venemans}, {Walter}, {Wang}, {Davies},
  {Decarli}, {Rosa}, {Drake}, {Fan}, {Mazzucchelli}, {Rix}, {Worseck}, \&
  {Yang}}]{Schindler2020}
{Schindler}, J.-T., {Farina}, E.~P., {Ba{\~n}ados}, E., {et~al.} 2020, \apj,
  905, 51

\bibitem[{{Simpson} {et~al.}(2014){Simpson}, {Mortlock}, {Warren}, {Cantalupo},
  {Hewett}, {McLure}, {McMahon}, \& {Venemans}}]{Simpson2014}
{Simpson}, C., {Mortlock}, D., {Warren}, S., {et~al.} 2014, \mnras, 442, 3454

\bibitem[{{Stiavelli} {et~al.}(2005){Stiavelli}, {Djorgovski}, {Pavlovsky},
  {Scarlata}, {Stern}, {Mahabal}, {Thompson}, {Dickinson}, {Panagia}, \&
  {Meylan}}]{Stiavelli2005}
{Stiavelli}, M., {Djorgovski}, S.~G., {Pavlovsky}, C., {et~al.} 2005, \apjl,
  622, L1

\bibitem[{{Thoul} \& {Weinberg}(1996)}]{ThoulWeinberg1996}
{Thoul}, A.~A. \& {Weinberg}, D.~H. 1996, \apj, 465, 608

\bibitem[{{Utsumi} {et~al.}(2010){Utsumi}, {Goto}, {Kashikawa}, {Miyazaki},
  {Komiyama}, {Furusawa}, \& {Overzier}}]{Utsumi2010}
{Utsumi}, Y., {Goto}, T., {Kashikawa}, N., {et~al.} 2010, \apj, 721, 1680

\bibitem[{{Valdes} {et~al.}(2014){Valdes}, {Gruendl}, \& {DES
  Project}}]{Valdes2014}
{Valdes}, F., {Gruendl}, R., \& {DES Project}. 2014, in Astronomical Society of
  the Pacific Conference Series, Vol. 485, Astronomical Data Analysis Software
  and Systems XXIII, ed. N.~{Manset} \& P.~{Forshay}, 379

\bibitem[{{Venemans} {et~al.}(2013){Venemans}, {Findlay}, {Sutherland}, {De
  Rosa}, {McMahon}, {Simcoe}, {Gonz{\'a}lez-Solares}, {Kuijken}, \&
  {Lewis}}]{Venemans2013}
{Venemans}, B.~P., {Findlay}, J.~R., {Sutherland}, W.~J., {et~al.} 2013, \apj,
  779, 24

\bibitem[{{Venemans} {et~al.}(2007){Venemans}, {McMahon}, {Warren},
  {Gonzalez-Solares}, {Hewett}, {Mortlock}, {Dye}, \& {Sharp}}]{Venemans2007}
{Venemans}, B.~P., {McMahon}, R.~G., {Warren}, S.~J., {et~al.} 2007, \mnras,
  376, L76

\bibitem[{{Venemans} {et~al.}(2016){Venemans}, {Walter}, {Zschaechner},
  {Decarli}, {De Rosa}, {Findlay}, {McMahon}, \& {Sutherland}}]{Venemans2016}
{Venemans}, B.~P., {Walter}, F., {Zschaechner}, L., {et~al.} 2016, \apj, 816,
  37

\bibitem[{{Verhamme} {et~al.}(2018){Verhamme}, {Garel}, {Ventou}, {Contini},
  {Bouch{\'e}}, {Herenz}, {Richard}, {Bacon}, {Schmidt}, {Maseda}, {Marino},
  {Brinchmann}, {Cantalupo}, {Caruana}, {Cl{\'e}ment}, {Diener}, {Drake},
  {Hashimoto}, {Inami}, {Kerutt}, {Kollatschny}, {Leclercq}, {Patr{\'\i}cio},
  {Schaye}, {Wisotzki}, \& {Zabl}}]{Verhamme2018}
{Verhamme}, A., {Garel}, T., {Ventou}, E., {et~al.} 2018, \mnras, 478, L60

\bibitem[{{Volonteri}(2012)}]{Volontari2012}
{Volonteri}, M. 2012, Science, 337, 544

\bibitem[{{Wang} {et~al.}(2021){Wang}, {Yang}, {Fan}, {Hennawi}, {Barth},
  {Banados}, {Bian}, {Boutsia}, {Connor}, {Davies}, {Decarli}, {Eilers},
  {Farina}, {Green}, {Jiang}, {Li}, {Mazzucchelli}, {Nanni}, {Schindler},
  {Venemans}, {Walter}, {Wu}, \& {Yue}}]{Wang2021}
{Wang}, F., {Yang}, J., {Fan}, X., {et~al.} 2021, \apjl, 907, L1

\bibitem[{{Wang} {et~al.}(2023){Wang}, {Yang}, {Hennawi}, {Fan}, {Sun},
  {Champagne}, {Costa}, {Habouzit}, {Endsley}, {Li}, {Lin}, {Meyer},
  {Schindler}, {Wu}, {Ba{\~n}ados}, {Barth}, {Bhowmick}, {Bieri}, {Blecha},
  {Bosman}, {Cai}, {Colina}, {Connor}, {Davies}, {Decarli}, {De Rosa}, {Drake},
  {Egami}, {Eilers}, {Evans}, {Farina}, {Haiman}, {Jiang}, {Jin}, {Jun},
  {Kakiichi}, {Khusanova}, {Kulkarni}, {Li}, {Liu}, {Loiacono}, {Lupi},
  {Mazzucchelli}, {Onoue}, {Pudoka}, {Rojas-Ruiz}, {Shen}, {Strauss}, {Tee},
  {Trakhtenbrot}, {Trebitsch}, {Venemans}, {Volonteri}, {Walter}, {Xie}, {Yue},
  {Zhang}, {Zhang}, \& {Zou}}]{Wang2023}
{Wang}, F., {Yang}, J., {Hennawi}, J.~F., {et~al.} 2023, \apjl, 951, L4

\bibitem[{{Willott} {et~al.}(2010){Willott}, {Delorme}, {Reyl{\'e}}, {Albert},
  {Bergeron}, {Crampton}, {Delfosse}, {Forveille}, {Hutchings}, {McLure},
  {Omont}, \& {Schade}}]{Willott2010}
{Willott}, C.~J., {Delorme}, P., {Reyl{\'e}}, C., {et~al.} 2010, \aj, 139, 906

\bibitem[{{Willott} {et~al.}(2005){Willott}, {Percival}, {McLure}, {Crampton},
  {Hutchings}, {Jarvis}, {Sawicki}, \& {Simard}}]{Willott2005}
{Willott}, C.~J., {Percival}, W.~J., {McLure}, R.~J., {et~al.} 2005, \apj, 626,
  657

\bibitem[{{Wold} {et~al.}(2019){Wold}, {Kawinwanichakij}, {Stevans},
  {Finkelstein}, {Papovich}, {Devarakonda}, {Ciardullo}, {Feldmeier}, {Florez},
  {Gawiser}, {Gronwall}, {Jogee}, {Marshall}, {Sherman}, {Shipley},
  {Somerville}, {Valdes}, \& {Zeimann}}]{Wold2019}
{Wold}, I. G.~B., {Kawinwanichakij}, L., {Stevans}, M.~L., {et~al.} 2019,
  \apjs, 240, 5

\bibitem[{{Wold} {et~al.}(2022){Wold}, {Malhotra}, {Rhoads}, {Wang}, {Hu},
  {Perez}, {Zheng}, {Khostovan}, {Walker}, {Barrientos},
  {Gonz{\'a}lez-L{\'o}pez}, {Harish}, {Infante}, {Jiang}, {Pharo},
  {Moya-Sierralta}, {Bauer}, {Galaz}, {Valdes}, \& {Yang}}]{Wold2022}
{Wold}, I. G.~B., {Malhotra}, S., {Rhoads}, J., {et~al.} 2022, \apj, 927, 36

\bibitem[{{Yang} {et~al.}(2023){Yang}, {Wang}, {Fan}, {Hennawi}, {Barth},
  {Ba{\~n}ados}, {Sun}, {Liu}, {Cai}, {Jiang}, {Li}, {Onoue}, {Schindler},
  {Shen}, {Wu}, {Bhowmick}, {Bieri}, {Blecha}, {Bosman}, {Champagne}, {Colina},
  {Connor}, {Costa}, {Davies}, {Decarli}, {De Rosa}, {Drake}, {Egami},
  {Eilers}, {Evans}, {Farina}, {Habouzit}, {Haiman}, {Jin}, {Jun}, {Kakiichi},
  {Khusanova}, {Kulkarni}, {Loiacono}, {Lupi}, {Mazzucchelli}, {Pan},
  {Rojas-Ruiz}, {Strauss}, {Tee}, {Trakhtenbrot}, {Trebitsch}, {Venemans},
  {Vestergaard}, {Volonteri}, {Walter}, {Xie}, {Yue}, {Zhang}, {Zhang}, \&
  {Zou}}]{Yang2023}
{Yang}, J., {Wang}, F., {Fan}, X., {et~al.} 2023, \apjl, 951, L5

\bibitem[{{Zewdie} {et~al.}(2023){Zewdie}, {Assef}, {Mazzucchelli}, {Aravena},
  {Blain}, {D{\'\i}az-Santos}, {Eisenhardt}, {Jun}, {Stern}, {Tsai}, \&
  {Wu}}]{Dejene2023}
{Zewdie}, D., {Assef}, R.~J., {Mazzucchelli}, C., {et~al.} 2023, \aap, 677, A54

\bibitem[{{Zheng} {et~al.}(2006){Zheng}, {Overzier}, {White}, {Ford},
  {Ben{\'\i}tez}, {Blakeslee}, {Bradley}, {Jee}, {Martel}, {Mei}, {Zirm},
  {Illingworth}, {Clampin}, {Hartig}, {Ardila}, {Bartko}, {Broadhurst},
  {Brown}, {Burrows}, {Cheng}, {Cross}, {Demarco}, {Feldman}, {Franx},
  {Golimowski}, {Goto}, {Gronwall}, {Holden}, {Homeier}, {Infante}, {Kimble},
  {Krist}, {Lesser}, {Menanteau}, {Meurer}, {Miley}, {Motta}, {Postman},
  {Rosati}, {Sirianni}, {Sparks}, {Tran}, \& {Tsvetanov}}]{Zheng2006}
{Zheng}, W., {Overzier}, R.~A. asnd~{Bouwens}, R.~J., {White}, R.~L., {et~al.}
  2006, \apj, 640, 574

\bibitem[{{Zheng} {et~al.}(2019){Zheng}, {Rhoads}, {Wang}, {Malhotra},
  {Walker}, {Mooney}, {Jiang}, {Hu}, {Hibon}, {Jiang}, {Infante}, {Barrientos},
  {Galaz}, {Valdes}, {Wester}, {Yang}, {Coughlin}, {Harish}, {Kang},
  {Khostovan}, {Kong}, {Perez}, {Pharo}, {Wold}, \& {Zheng}}]{Zheng2019}
{Zheng}, Z.-Y., {Rhoads}, J.~E., {Wang}, J.-X., {et~al.} 2019, \pasp, 131,
  074502

\bibitem[{{Zheng} {et~al.}(2017){Zheng}, {Wang}, {Rhoads}, {Infante},
  {Malhotra}, {Hu}, {Walker}, {Jiang}, {Jiang}, {Hibon}, {Gonzalez}, {Kong},
  {Zheng}, {Galaz}, \& {Barrientos}}]{Zheng2017}
{Zheng}, Z.-Y., {Wang}, J., {Rhoads}, J., {et~al.} 2017, \apjl, 842, L22

\bibitem[{{Zhou} {et~al.}(2023){Zhou}, {Chen}, {Di Matteo}, {Ni}, {Croft}, \&
  {Bird}}]{Zhou2023}
{Zhou}, Y., {Chen}, H., {Di Matteo}, T., {et~al.} 2023, arXiv e-prints,
  arXiv:2309.11571

\end{thebibliography}

\begin{appendix}
    \section{Additional Material}
\begin{table*}[!htb]
%\vspace{10cm}
\caption{Names, Coordinates, projected distances from the quasar, i-band, z-band, and NB964-band magnitudes, Ly-$\alpha$ luminosities, and star formation rates for the LAE candidates identified in this work}
\resizebox{\linewidth}{!}{
\begin{tabular}{cllcrrlllrr}
\hline \hline \\
\multicolumn{1}{c}{ID} & \multicolumn{1}{c}{R.A.}       & \multicolumn{1}{c}{Dec.}          & \multicolumn{1}{c}{$\theta$} & \multicolumn{1}{c}{$r_c$} & \multicolumn{1}{c}{$r_p$} & \multicolumn{1}{c}{i} & \multicolumn{1}{c}{z} & \multicolumn{1}{c}{NB964}   & \multicolumn{1}{c}{$L_{\text{Ly}\alpha}$}            & \multicolumn{1}{c}{SFR}                     \\\\ 
   & \multicolumn{1}{c}{(J2000.00)} & \multicolumn{1}{c}{(J2000.00)}    & \multicolumn{1}{c}{(deg)}            & \multicolumn{1}{c}{(cMpc)}            & \multicolumn{1}{c}{(pMpc)}          &    \multicolumn{1}{c}{}  &     \multicolumn{1}{c}{} &  \multicolumn{1}{c}{}        & \multicolumn{1}{c}{$\times 10^{42}$ (ergs s$^{-1}$)} & \multicolumn{1}{c}{($M_{\odot}$ yr$^{-1}$)} \\\hline
QSO & 23:48:33.33 & -30:54:10.23 & 0.00 & 0.00   & 0.00  & \textgreater{}27.15 & 22.19 $\pm$ 0.05  & 20.86 $\pm$ 0.01 & - & - \\
LAE-1  & 23:50:39.46 & -31:47:26.41 & 0.99 & 149.20 & 18.89 & \textgreater{}27.15 & 24.97 $\pm$ 0.16  & 24.11 $\pm$ 0.12 & 7.6 $\pm$ 2.7   & 4.7 $\pm$ 1.7   \\
LAE-2  & 23:49:25.39 & -31:46:34.95 & 0.89 & 133.93 & 16.95 & \textgreater{}27.15 & 26.07 $\pm$ 0.43  & 24.73 $\pm$ 0.22 & 6.6 $\pm$ 2.7   & 4.1 $\pm$ 1.7   \\
LAE-3  & 23:49:45.53 & -31:39:01.46 & 0.79 & 118.57 & 15.01 & \textgreater{}27.15 & 25.74 $\pm$ 0.33  & 24.59 $\pm$ 0.19 & 6.6 $\pm$ 2.8   & 4.0 $\pm$ 1.7   \\
LAE-4  & 23:48:22.23 & -31:37:53.92 & 0.73 & 109.47 & 13.86 & \textgreater{}27.15 & 25.93 $\pm$ 0.38  & 24.46 $\pm$ 0.17 & 9.0 $\pm$ 2.7   & 5.5 $\pm$ 1.7   \\
LAE-5  & 23:48:17.08 & -31:30:18.94 & 0.61 & 90.77  & 11.49 & \textgreater{}27.15 & 26.09 $\pm$ 0.45  & 23.59 $\pm$ 0.07 & 26.5 $\pm$ 2.7  & 16.4 $\pm$ 1.7  \\
LAE-6  & 23:46:51.65 & -31:19:29.93 & 0.56 & 83.48  & 10.57 & \textgreater{}27.15 & 25.85 $\pm$ 0.36  & 24.37 $\pm$ 0.16 & 9.8 $\pm$ 2.8   & 6.1 $\pm$ 1.7   \\
LAE-7  & 23:52:04.95 & -31:16:20.60 & 0.84 & 126.08 & 15.96 & \textgreater{}27.15 & 25.38 $\pm$ 0.24  & 23.50 $\pm$ 0.07 & 25.5 $\pm$ 2.8  & 15.7 $\pm$ 1.7  \\
LAE-8  & 23:44:51.34 & -31:14:01.78 & 0.86 & 128.78 & 16.30 & \textgreater{}27.15 & 25.99 $\pm$ 0.42  & 24.08 $\pm$ 0.12 & 15.0 $\pm$ 2.8  & 9.2 $\pm$ 1.7   \\
LAE-9  & 23:49:09.11 & -31:11:23.68 & 0.31 & 47.13  & 5.97  & \textgreater{}27.15 & 25.84 $\pm$ 0.36  & 24.53 $\pm$ 0.18 & 7.7 $\pm$ 2.8   & 4.8 $\pm$ 1.7   \\
LAE-10 & 23:48:38.18 & -31:10:25.49 & 0.27 & 40.72  & 5.15  & \textgreater{}27.15 & 26.12 $\pm$ 0.47  & 24.49 $\pm$ 0.18 & 9.4 $\pm$ 2.8   & 5.8 $\pm$ 1.7   \\
LAE-11 & 23:52:11.44 & -31:10:09.82 & 0.82 & 123.43 & 15.62 & \textgreater{}27.15 & 25.55 $\pm$ 0.28  & 24.47 $\pm$ 0.18 & 6.9 $\pm$ 2.8   & 4.2 $\pm$ 1.8   \\
LAE-12 & 23:50:09.67 & -31:00:28.43 & 0.36 & 53.98  & 6.83  & \textgreater{}27.15 & 25.54 $\pm$ 0.28  & 24.20 $\pm$ 0.14 & 10.7 $\pm$ 2.8  & 6.6 $\pm$ 1.7   \\
LAE-13 & 23:52:50.08 & -30:59:23.05 & 0.92 & 138.21 & 17.50 & \textgreater{}27.15 & 25.81 $\pm$ 0.36  & 24.37 $\pm$ 0.16 & 9.7 $\pm$ 2.9   & 6.0 $\pm$ 1.8   \\
LAE-14 & 23:46:52.33 & -30:57:07.02 & 0.36 & 54.65  & 6.92  & \textgreater{}27.15 & 26.07 $\pm$ 0.44  & 24.66 $\pm$ 0.22 & 7.3 $\pm$ 2.8   & 4.5 $\pm$ 1.7   \\
LAE-15 & 23:52:39.27 & -30:51:50.73 & 0.88 & 132.02 & 16.71 & \textgreater{}27.15 & 25.44 $\pm$ 0.27 & 24.32 $\pm$ 0.16 & 8.1 $\pm$ 2.9   & 5.0 $\pm$ 1.8   \\
LAE-16 & 23:51:42.32 & -30:50:46.97 & 0.68 & 101.71 & 12.87 & \textgreater{}27.15 & 25.33 $\pm$ 0.23 & 24.46 $\pm$ 0.16 & 5.5 $\pm$ 2.7   & 3.4 $\pm$ 1.7   \\
LAE-17 & 23:45:30.76 & -30:50:42.20 & 0.66 & 98.31  & 12.44 & \textgreater{}27.15 & 25.86 $\pm$ 0.37  & 24.72 $\pm$ 0.22 & 5.8 $\pm$ 2.8   & 3.6 $\pm$ 1.7   \\
LAE-18 & 23:45:38.56 & -30:41:06.23 & 0.66 & 99.34  & 12.57 & \textgreater{}27.15 & 25.69 $\pm$ 0.32  & 24.28 $\pm$ 0.14 & 10.4 $\pm$ 2.7  & 6.4 $\pm$ 1.7   \\
LAE-19 & 23:50:56.96 & -30:40:16.95 & 0.56 & 84.55  & 10.70 & \textgreater{}27.15 & 26.03 $\pm$ 0.43  & 24.07 $\pm$ 0.12 & 15.3 $\pm$ 2.8  & 9.5 $\pm$ 1.7   \\
LAE-20 & 23:46:20.25 & -30:40:03.94 & 0.53 & 79.67  & 10.08 & \textgreater{}27.15 & 25.38 $\pm$ 0.24  & 24.57 $\pm$ 0.19 & 4.6 $\pm$ 2.8   & 2.8 $\pm$ 1.7   \\
LAE-21 & 23:48:12.27 & -30:32:34.32 & 0.37 & 55.15  & 6.98  & \textgreater{}27.15 & 26.02 $\pm$ 0.42  & 24.22 $\pm$ 0.14 & 12.9 $\pm$ 2.8  & 7.9 $\pm$ 1.7   \\
LAE-22 & 23:44:45.07 & -30:31:07.12 & 0.90 & 135.50 & 17.15 & \textgreater{}27.15 & 26.05 $\pm$ 0.45  & 24.52 $\pm$ 0.18 & 8.8 $\pm$ 2.8   & 5.4 $\pm$ 1.7   \\
LAE-23 & 23:47:37.50 & -30:28:28.35 & 0.47 & 70.89  & 8.97  & \textgreater{}27.15 & 25.53 $\pm$ 0.27  & 24.60 $\pm$ 0.20 & 5.2 $\pm$ 2.8   & 3.2 $\pm$ 1.7   \\
LAE-24 & 23:44:31.64 & -30:27:35.79 & 0.97 & 145.88 & 18.47 & \textgreater{}27.15 & 25.39 $\pm$ 0.24  & 24.41 $\pm$ 0.17 & 6.6 $\pm$ 2.8   & 4.1 $\pm$ 1.8   \\
LAE-25 & 23:47:38.55 & -30:25:47.75 & 0.51 & 76.79  & 9.72  & \textgreater{}27.15 & 25.33 $\pm$ 0.23  & 24.49 $\pm$ 0.18 & 5.1 $\pm$ 2.8   & 3.2 $\pm$ 1.7   \\
LAE-26 & 23:44:36.91 & -30:25:04.43 & 0.98 & 146.42 & 18.53 & \textgreater{}27.15 & 25.95 $\pm$ 0.40  & 23.97 $\pm$ 0.11 & 17.0 $\pm$ 2.8  & 10.5 $\pm$ 1.7  \\
LAE-27 & 23:48:21.02 & -30:24:04.19 & 0.50 & 75.52  & 9.56  & \textgreater{}27.15 & 26.00 $\pm$ 0.41  & 24.52 $\pm$ 0.18 & 8.5 $\pm$ 2.8   & 5.3 $\pm$ 1.7   \\
LAE-28 & 23:47:21.19 & -30:22:53.54 & 0.58 & 87.27  & 11.05 & \textgreater{}27.15 & 25.63 $\pm$ 0.30  & 23.84 $\pm$ 0.10 & 18.1 $\pm$ 2.8  & 11.2 $\pm$ 1.7  \\
LAE-29 & 23:46:27.28 & -30:21:51.57 & 0.70 & 105.43 & 13.35 & \textgreater{}27.15 & 25.52 $\pm$ 0.27  & 24.40 $\pm$ 0.16 & 7.6 $\pm$ 2.8   & 4.7 $\pm$ 1.7   \\
LAE-30 & 23:49:27.05 & -30:21:25.98 & 0.58 & 86.77  & 10.98 & \textgreater{}27.15 & 25.65 $\pm$ 0.30  & 24.61 $\pm$ 0.19 & 5.9 $\pm$ 2.7   & 3.6 $\pm$ 1.7   \\
LAE-31 & 23:50:27.20 & -30:21:13.39 & 0.68 & 102.61 & 12.99 & \textgreater{}27.15 & 25.50 $\pm$ 0.27  & 24.26 $\pm$ 0.14 & 9.5 $\pm$ 2.8   & 5.8 $\pm$ 1.7   \\
LAE-32 & 23:45:02.91 & -30:12:32.87 & 1.03 & 153.76 & 19.46 & \textgreater{}27.15 & 25.96 $\pm$ 0.39  & 24.73 $\pm$ 0.21 & 6.1 $\pm$ 2.7   & 3.8 $\pm$ 1.7   \\
LAE-33 & 23:49:34.65 & -30:12:40.49 & 0.73 & 108.83 & 13.78 & \textgreater{}27.15 & 25.66 $\pm$ 0.31  & 24.52 $\pm$ 0.18 & 6.9 $\pm$ 2.8   & 4.3 $\pm$ 1.7   \\
LAE-34 & 23:50:23.24 & -30:10:45.41 & 0.82 & 123.58 & 15.64 & \textgreater{}27.15 & 26.13 $\pm$ 0.48  & 24.62 $\pm$ 0.20 & 7.9 $\pm$ 2.8   & 4.9 $\pm$ 1.7   \\
LAE-35 & 23:50:45.07 & -30:05:13.35 & 0.94 & 141.41 & 17.90 & \textgreater{}27.15 & 25.34 $\pm$ 0.23  & 24.32 $\pm$ 0.15 & 7.5 $\pm$ 2.8   & 4.6 $\pm$ 1.7   \\
LAE-36 & 23:49:43.48 & -30:03:01.26 & 0.89 & 133.31 & 16.87 & \textgreater{}27.15 & 25.55 $\pm$ 0.28  & 24.31 $\pm$ 0.15 & 9.1 $\pm$ 2.8   & 5.6 $\pm$ 1.7   \\
LAE-37 & 23:50:36.54 & -30:01:46.79 & 0.98 & 146.80 & 18.58 & \textgreater{}27.15 & 25.45 $\pm$ 0.25  & 24.58 $\pm$ 0.19 & 4.9 $\pm$ 2.8   & 3.0 $\pm$ 1.7   \\
LAE-38 & 23:48:43.58 & -29:53:13.94 & 1.02 & 152.41 & 19.29 & \textgreater{}27.15 & 25.80 $\pm$ 0.36  & 24.70 $\pm$ 0.22 & 5.7 $\pm$ 2.9   & 3.5 $\pm$ 1.8  
\\ \hline
\end{tabular}}
\label{tbl:catalogue}
\end{table*}

\newpage \ \newpage

\begin{figure*}
\includegraphics[width=0.5\textwidth]{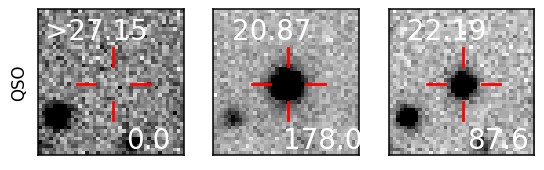}
\includegraphics[width=0.5\textwidth]{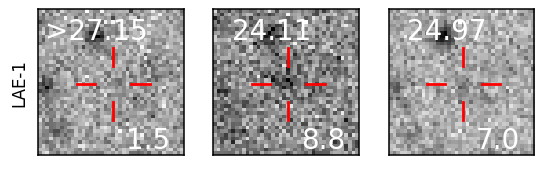}
\includegraphics[width=0.5\textwidth]{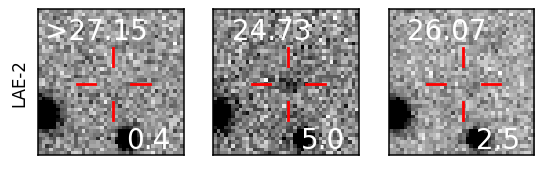}
\includegraphics[width=0.5\textwidth]{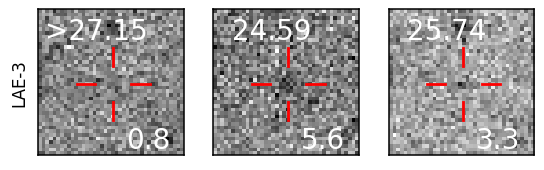}
\includegraphics[width=0.5\textwidth]{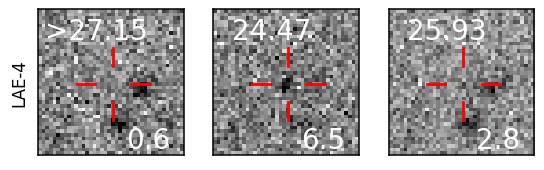}
\includegraphics[width=0.5\textwidth]{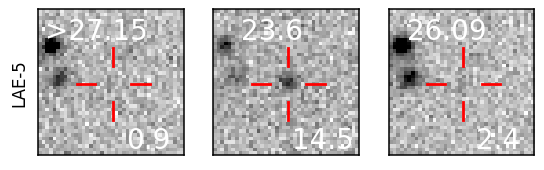}
\includegraphics[width=0.5\textwidth]{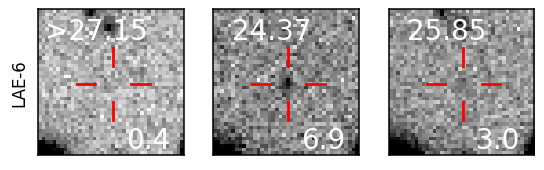}
\includegraphics[width=0.5\textwidth]{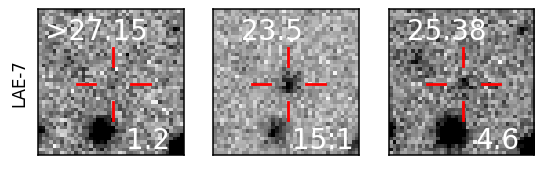}
\includegraphics[width=0.5\textwidth]{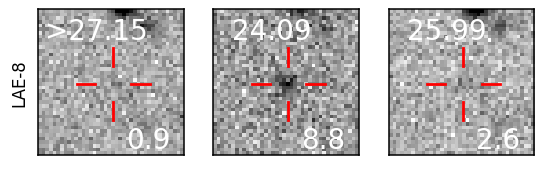}
\includegraphics[width=0.5\textwidth]{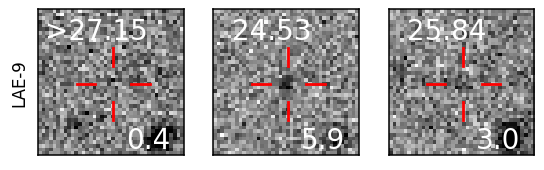}
\includegraphics[width=0.5\textwidth]{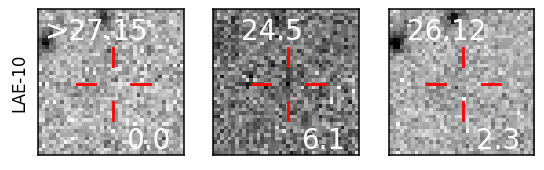}
\includegraphics[width=0.5\textwidth]{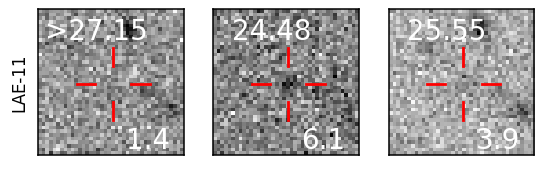}
\includegraphics[width=0.5\textwidth]{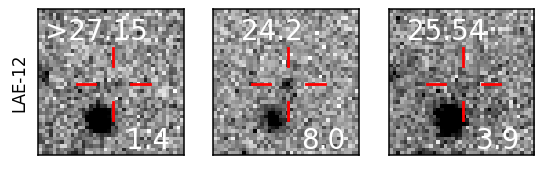}
\includegraphics[width=0.5\textwidth]{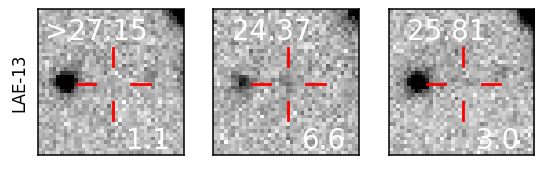}
\includegraphics[width=0.5\textwidth]{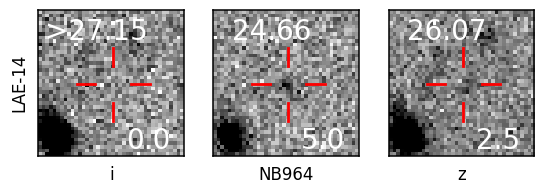}
\includegraphics[width=0.5\textwidth]{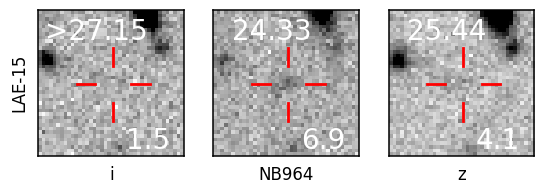}
\caption{LAE candidate 20 pix $\times$ 20 pix (10.8" $\times$ 10.8") cutouts, identified in our DECam data. The values in the top of each postage stamp indicate the AB magnitudes in the i, NB964, and z bands whilst the bottom show the S/N values in those respective bands.}
\label{fig: LAE-candidates}
\end{figure*}

\renewcommand{\thefigure}{\thesection.\arabic{figure} (Cont.)}
\addtocounter{figure}{-1}

\begin{figure*}
\includegraphics[width=0.5\textwidth]{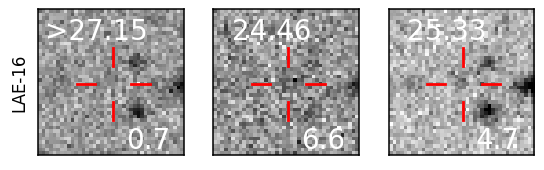}
\includegraphics[width=0.5\textwidth]{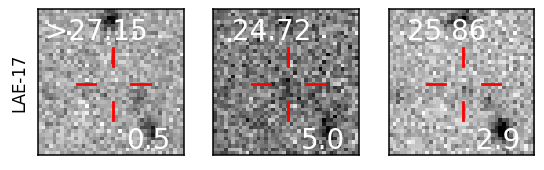}
\includegraphics[width=0.5\textwidth]{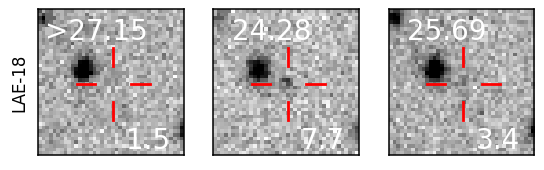}
\includegraphics[width=0.5\textwidth]{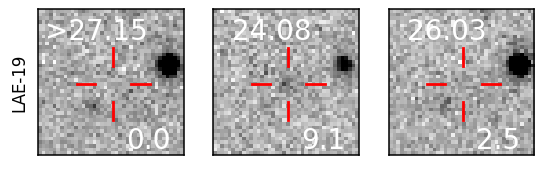}
\includegraphics[width=0.5\textwidth]{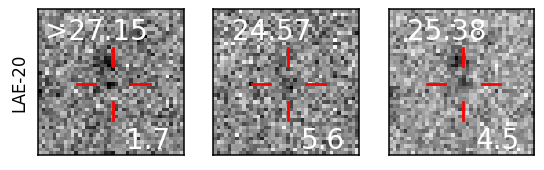}
\includegraphics[width=0.5\textwidth]{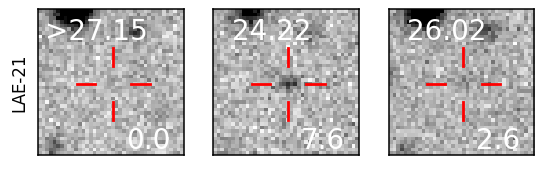}
\includegraphics[width=0.5\textwidth]{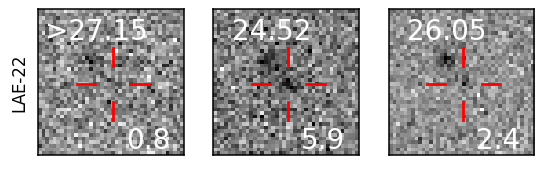}
\includegraphics[width=0.5\textwidth]{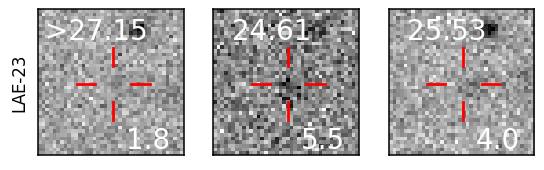}
\includegraphics[width=0.5\textwidth]{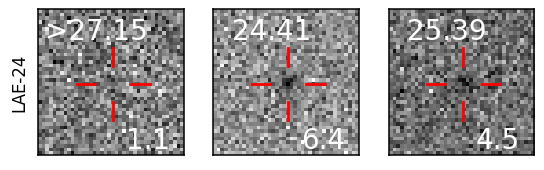}
\includegraphics[width=0.5\textwidth]{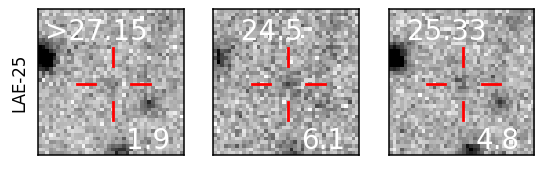}
\includegraphics[width=0.5\textwidth]{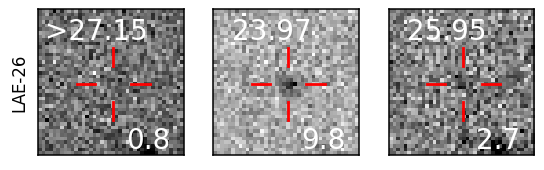}
\includegraphics[width=0.5\textwidth]{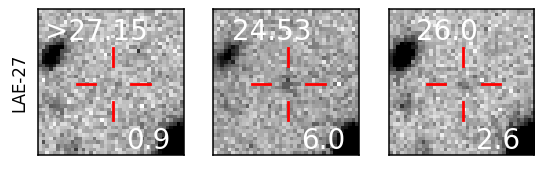}
\includegraphics[width=0.5\textwidth]{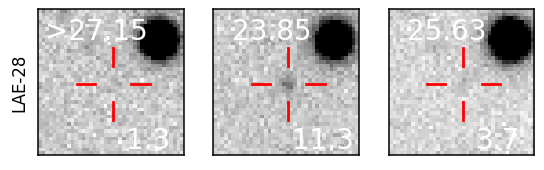}
\includegraphics[width=0.5\textwidth]{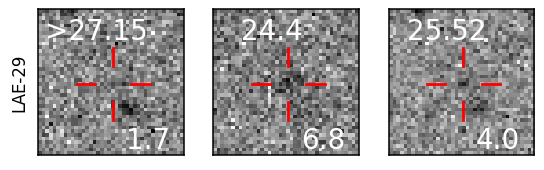}
\includegraphics[width=0.5\textwidth]{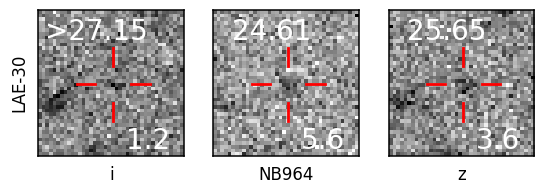}
\includegraphics[width=0.5\textwidth]{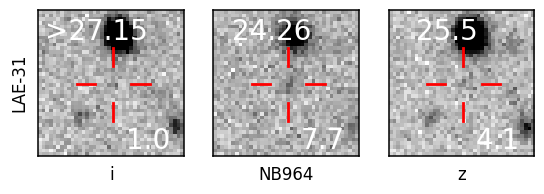}
\caption{}
\end{figure*}

\renewcommand{\thefigure}{\thesection.\arabic{figure} (Cont.)}
\addtocounter{figure}{-1}

\begin{figure}
\includegraphics[width=0.5\textwidth]{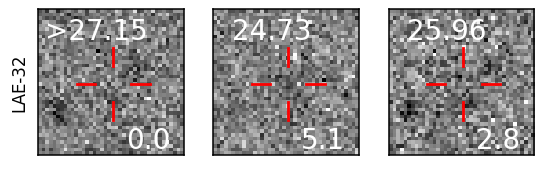}
\includegraphics[width=0.5\textwidth]{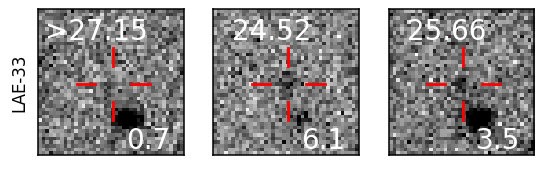}
\includegraphics[width=0.5\textwidth]{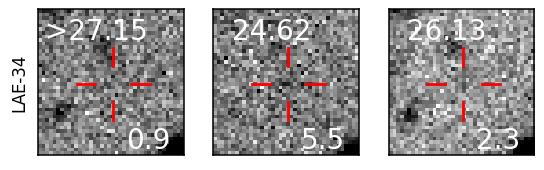}
\includegraphics[width=0.5\textwidth]{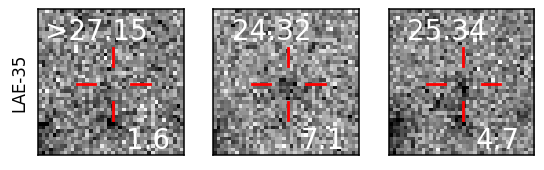}
\includegraphics[width=0.5\textwidth]{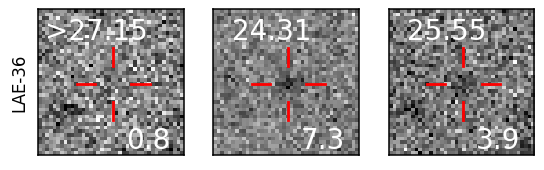}
\includegraphics[width=0.5\textwidth]{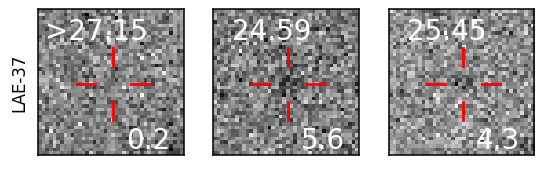}
\includegraphics[width=0.5\textwidth]{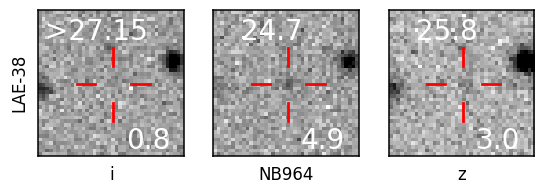}
\caption{}
\end{figure}

\renewcommand{\thefigure}{\arabic{figure}}

\end{appendix}
\end{document}